%% file: 0.main.tex
\documentclass[manuscript,review=false,anonymous=false]{acmart} 

\AtBeginDocument{%
  \providecommand\BibTeX{{%
    \normalfont B\kern-0.5em{\scshape i\kern-0.25em b}\kern-0.8em\TeX}}}

\setcopyright{acmcopyright}
\copyrightyear{2018}
\acmYear{2018}
\acmDOI{10.1145/1122445.1122456}

\usepackage{booktabs} 
\usepackage{url}

\usepackage{caption}
\usepackage[all]{nowidow}
\usepackage{wrapfig}
\usepackage{array}
\usepackage{arydshln}
\usepackage{tabularx}
\usepackage{multirow}
\usepackage{arydshln}
\newcolumntype{L}[1]{>{\raggedright\let\newline\\\arraybackslash\hspace{0pt}}m{#1}}
\newcolumntype{C}[1]{>{\centering\let\newline\\\arraybackslash\hspace{0pt}}m{#1}}
\newcolumntype{R}[1]{>{\raggedleft\let\newline\\\arraybackslash\hspace{0pt}}m{#1}}

\usepackage{wrapfig,lipsum,booktabs} 

\def\authnotes{1}
\newcounter{notectr}[section]
\newcommand{\thenote}{\thesubsection.\arabic{notectr}\refstepcounter{notectr}}

\newcommand{\note}[2]{$\ll$#1~\thenote: #2$\gg$}
\newcommand{\cnote}[1]{\ifnum\authnotes=1 \textcolor{blue}{\note{Comment:}{#1}}\fi}

\copyrightyear{2026}
\acmYear{2026}
\setcopyright{acmlicensed}\acmConference[CHI '26]{CHI}{Nov 5-9, 2026}{USA}
\acmBooktitle{CHI'24, Nov, 2026, USA}
\acmPrice{15.00}
\acmDOI{10.1145/3491102.43}
\acmISBN{978-1-4503-9157-3/22/04}

\begin{document}



\title[Emotional Support AI]{Emotional Support with Conversational AI: Talking to Machines About Life}


\author{Olivia Yan Huang}
\affiliation{%
  \institution{University of Illinois Urbana-Champaign}
  \country{USA}}
\email{yan65@illinois.edu}

\author{Monika Stodolska}
\affiliation{%
  \institution{University of Illinois Urbana-Champaign}
  \country{USA}}
\email{stodolsk@illinois.edu}

\author{Sharifa Sultana}
\affiliation{%
  \institution{University of Illinois Urbana-Champaign}
  \country{USA}}
\email{sharifas@illinois.edu}

\renewcommand{\shortauthors}{Huang et al.}

\begin{abstract}
AI companion chatbots are increasingly used for emotional support, with prior work in the domain predominantly documenting their mixed psychosocial impacts, including both increased emotional expression and heightened loneliness. However, most among existing research primarily focus on outcome-level effects, offering limited insight into how emotional support is produced through interaction. In this paper, we examine emotional support as an interactional and socially situated process. Drawing on qualitative analysis of Reddit discussions, we analyze how users engage with AI companions and how these interactions are interpreted and contested within online communities. We show that emotional support is co-constructed through conversational mechanisms such as validation, reflective prompting, and companionship, while also giving rise to tensions including support versus dependency, validation versus delusion, and accessibility versus harm. Importantly, support extends beyond human–AI interaction and is shaped by community responses that legitimize or challenge AI-mediated care. Hence, we reconceptualize AI emotional support as a negotiated socio-technical process and derive implications for designing responsible- and context-sensitive AI systems.
\end{abstract}







\settopmatter{printfolios=true}

\maketitle

\input{1.intro}
\input{2.lit}
\input{4.context}
\input{3.methods}

\input{5.find} 
\input{6.discussion}
\input{6.lim-fw-con}



\bibliographystyle{ACM-Reference-Format}
\bibliography{0.main}

\end{document}

%% file: 1.intro.tex
\section{Introduction}
Large language models (LLMs) are increasingly embedded in everyday life as conversational agents capable of producing fluent dialogue and responding to complex questions. While early discussions of LLMs focused primarily on their capabilities for information retrieval, programming assistance, or productivity tasks \cite{haleem2022era, alzu2024exploring}, emerging evidence suggests that users are also turning to these systems for deeply personal interactions \cite{delessons, song2022can, xie2023friend134, loveys2022felt, wang2025my}. Across online communities like Reddit, people describe talking to LLMs about loneliness, emotional distress, relationship problems, and even existential questions about life meaning and purpose. These interactions suggest that conversational AI systems are beginning to play a role in how individuals seek and obtain emotional support and guidance.

Emotional support has traditionally been situated within human relationships and institutional structures such as friendships, family networks, therapy, or religious communities. However, LLMs offer a distinct set of socio-technical affordances that reshape how such support can be accessed. They are persistently available \cite{xygkou2023conversation}, anonymous, non-judgmental \cite{zhang2025qualitative}, and free from the interpersonal obligations that often accompany emotional disclosure in human relationships. As a result, individuals may turn to LLMs when they feel unable or unwilling to seek support from other people. Despite growing public attention to AI companionship and mental health chatbots \cite{hernandez2021smart, ho2025potential, denecke2021artificial, casu2024ai}, we still know relatively little about how people actually use general-purpose LLMs to navigate emotional challenges and questions of meaning in everyday life.

In this study, we investigate how people seek emotional support through interactions with LLMs. Drawing on qualitative analysis of Reddit discussions, we examine what qualities of LLMs make them attractive as sources of emotional support, what emotional needs they fulfill for people seeking emotional support, and the concerns that arise when AI systems mediate emotionally sensitive interactions. To interpret these practices, we draw on self-determination theory, which conceptualizes emotional well-being through three core psychological needs: relatedness, competence, and autonomy \cite{ryan2024self}. These needs correspond to users seeking companionship and intimacy, coping strategies and psychological guidance, and reflection on moral, existential, or spiritual questions.

Guided by this perspective, our study addresses the following research questions:

\begin{quote}
\textit{\textbf{RQ1:}} What affordances of LLMs lead people to seek emotional support from them rather than from human interlocutors? \\
\textit{\textbf{RQ2:}} How do interactions with LLMs address users’ needs for relatedness, competence, and autonomy? \\
\textit{\textbf{RQ3:}} What benefits and risks emerge when LLMs become sources of companionship, psychological coping, and existential reflection?
\end{quote}

\paragraph{Contributions}
This paper makes four contributions to research on human–AI interaction and emotional support technologies.

\begin{enumerate}
\item \textbf{Revealing what qualities of LLMs make them sought after by people as sources of emotional support.}
Through qualitative analysis of Reddit discussions, we identify socio-technical affordances that make LLMs attractive sources of emotional support, including persistent availability, anonymity, non-judgmental responses, and the absence of interpersonal burden. These affordances enable users to disclose vulnerability, reveal frustrations, and explore personal concerns in ways that may feel difficult within human relationships.

\item \textbf{Introducing a psychological framework for understanding emotional support practices with LLMs.}
Drawing on self-determination theory, we conceptualize emotional support interactions with LLMs across three psychological needs: \emph{relatedness} (companionship and intimacy), \emph{competence} (coping and psychological guidance), and \emph{autonomy} (existential and spiritual reflection). This framework provides a structured lens for understanding the diverse ways users seek emotional support through conversational AI.

\item \textbf{Conceptualizing LLMs as emerging emotional support infrastructure.}
Our findings show that LLMs are increasingly appropriated as informal infrastructures for emotional support that operate alongside, and sometimes in place of, traditional support systems such as friends, therapists, or religious communities. Users turn to LLMs to process emotions, rehearse difficult conversations, and reflect on personal beliefs.

\item \textbf{Exposing critical risks when LLMs mediate psychological and existential support.}
Although many users report benefits from AI-mediated emotional support, our analysis also reveals important tensions, including emotional dependency, unreliable mental health advice, crossing moral and ethical boundaries of gender and sexual relations, and the possibility of AI systems replacing existing moral or spiritual authorities. These findings highlight emerging ethical challenges when AI systems become embedded in domains traditionally shaped by human care and professional expertise.
\end{enumerate}

\textbf{Ethics, Disclosure, and Reflexivity.} We recognize the sensitivity and ethical concerns in this study and include a self-reflective disclosure on our work. (a) As our data are publicly accessible from Reddit, this work did not require institutional review board approval. We recognize that even though Reddit posts are publicly accessible, public visibility does not erase participants' contextual expectations of privacy \cite{zimmer2020but}. Our approach aligns with HCI calls to consider obligations not only to individuals but also to digital communities as a whole, particularly when studying vulnerable populations in sensitive contexts \cite{fiesler2016exploring}. Moreover, as prior research on chatrooms shows, participants often object to being studied without consent \cite{hudson2004go}, and we therefore removed personally identifiable information, paraphrased quoted content, and anonymized all excerpts. In doing so, we treated online discussions about emotional support, mental health struggles, and existential reflection as sensitive socio-technical contexts rather than neutral datasets.

(b) Our interpretations are informed by our perspectives as researchers studying human–AI interaction and the social implications of emerging technologies. While examining Reddit discussions about emotional support with large language models (LLMs), we remained attentive to the fact that users’ experiences with loneliness, psychological distress, and existential struggles are deeply personal and context-dependent. Our analysis therefore emphasizes patterns within the dataset while acknowledging that alternative interpretations are possible. In addition, considering that our study was grounded within an interpretivist paradigm, its goal was not generalizability of results, but rather presenting the unique perspectives of people sharing their experiences on Reddit. Similarly, Reddit is a forum-style platform that emphasizes open conversation and knowledge sharing on a variety of topics, and the views of people who post there may not be representative of other AI users.

(c) We clarify that our work is not intended to promote or normalize reliance on AI systems for emotional, psychological, or spiritual guidance, nor to reinforce stereotypes about people who seek support through online technologies. Instead, our goal is to examine how individuals describe using LLMs in moments of loneliness, distress, or reflection, and to critically analyze both the opportunities and risks that emerge when conversational AI becomes embedded in practices of emotional support. Despite our efforts, we acknowledge the potential ethical implications of studying sensitive discussions about emotional well-being and elaborate further on these issues in the Discussion. This paper reports on how people describe seeking emotional support from LLMs; we are not qualified to evaluate the accuracy of AI-generated advice and we do not advocate for AI to replace professional mental health or spiritual guidance.

\paragraph{Paper Structure}
The remainder of the paper is organized as follows. We first review prior work on emotional support technologies, conversational AI, and human–AI companionship. We then describe our dataset and qualitative analysis methods. Next, we present our findings across three dimensions of emotional support practices with LLMs: relatedness, competence, and autonomy. Finally, we discuss the implications of LLMs as emerging emotional support infrastructures, highlighting both opportunities and risks for the design and governance of conversational AI systems.

%% file: 2.lit.tex
\section{Related Work}

\subsection{Human-AI Companionship and Social Relationships with Conversational Agents} 
A growing body of literature has examined LLMs and conversational AI as a form of social or relational technology. AI companion chatbots and related social agents have been positioned as emotionally responsive entities capable of simulating social presence and relational interaction \cite{brandtzaeg2022my012, alabed2024more002, alotaibi2024the003, andersson2025companionship005}. As a result, people increasingly turn to LLMs and AI companions for emotional support, companionship, and even intimacy \cite{depounti2023ideal031, kim2024what056, folk2025individual039}. Prior studies have revealed that users may perceive such systems as friends, confidants, or companions, and that interactions with conversational agents can facilitate emotional disclosure, reflection, and perceived social connection \cite{dosovitsky2021bonding032, skjuve2021my113}. Such interactions are stimulated by the chatbots’ anthropomorphic design \cite{araujo2018living}, including human-like features, avatars, and physical and verbal embodiment cues that could be customized to one’s physical ideal \cite {bergner2023machine, bertacchini2017shopping, holzwarth2006influence, zierau2023voice}. AI chatbots also appear emotionally available, non-judgmental, and comforting\cite{xygkou2023conversation, zhang2025qualitative, zhou2023comforting}, while offering attentiveness \cite{ta2020user}, real-time emotional support \cite{ramadan2021amazon}, and empathy \cite{park2023generative}. LLM interactions fulfill two innate attachment needs that humans instinctively seek: comfort and emotional security \cite{rever1972attachment}. They may also elicit all three components of the triadic love structure \cite{sternberg1986triangular}, including intimacy, passion, and commitment \cite{skjuve2023longitudinal, alabed2024more, siemon2022we}. Prior studies have also found that AI companions can serve as readily available sources of perceived social support, helping to provide entertainment, learn people’s skills, and enhance well-being \cite{li2024finding, pan2024desirable}. By fulfilling these needs, they can relieve feelings of loneliness and distress \cite{depounti2023ideal, laestadius2024too, pan2024desirable, wygnanska2023experience, xie2022attachment}.

While LLMs and conversational AI offer many benefits, scholars have begun to highlight tensions and risks involved in human-AI interactions. Companion chatbots’ highly personalized, emotionally responsive, and engagement-driven design increases the risk of emotional overdependence, overreliance, addiction, and unrealistic expectations \cite{xie2023friend134, laestadius2024too061, chandra2025from018}. A key concern has also been raised with respect to AI companions’ ability to simulate intimacy without true reciprocity. When users begin perceiving chatbots as sentient partners with genuine emotional needs, this perception generates an illusory sense of mutual obligation that closely resembles unhealthy patterns of relational dependence \cite{laestadius2024too, pentina2023exploring}. With greater investment of time and emotions, these artificial relationships risk distorting users’ expectations of intimacy, straining or displacing authentic human bonds, and promoting social withdrawal along with a general weakening of interpersonal connections \cite{zhang2025qualitative}. Studies have also documented concerns related to commodification of users’ emotional labor, data privacy, and the fragility of relationships formed with systems that may change or disappear \cite{pan2025grooming085, banks2024deletion009, freitas2024lessons026}. In addition, AI companions may be exploited by companies or malicious actors who leverage intimacy to manipulate, spread misinformation, or coerce \cite{rosenberg2023manipulation, shank2025artificial}. Such threats may be amplified among socially vulnerable users, such as teenagers and people with mental health issues, and contribute to anxiety, depression, diminished well-being, and, in extreme cases, psychological distress and suicide \cite{castillo2024dating}. Broader social and ethical concerns have also been raised with respect to stigma associated with human-AI relationships \cite{smith2025can}, reinforcement of gender-based hostility \cite{leshner2024technically}, and wider sociocultural consequences if AI is used as a substitute for human intimacy \cite{cave2019hopes}. 

Despite these insights, existing work has largely focused on dedicated AI companion platforms or romantic chatbot systems. Less attention has been paid to how general-purpose LLMs become embedded in everyday emotional practices beyond providing companionship. Moreover, prior research has primarily examined relational attachments, while comparatively little work has explored how conversational AI may also function as a space for psychological guidance, moral reasoning, or spiritual reflection. Our findings extend this literature by showing that users engage LLMs not only as companions but also as conversational partners helpful in navigating emotional distress, developing coping strategies, and answering existential questions. In addition, to the best of our knowledge, none of the studies to date have examined the role of LLMs in the fulfillment of users’ basic needs through the lens of self-determination theory \cite{deci2012self}. By analyzing Reddit discussions across communities focused on AI, mental health, and religion, our study examines how LLM interactions support needs for relatedness, competence, and autonomy (RQ2) and explores the benefits and risks that emerge when AI systems become sources of emotional and existential support (RQ3).

\subsection{Emotional Support Technologies and Mental Health Chatbots}
Chatbots have long been studied as conversational systems that support information access, task completion, and interpersonal interaction \cite{benke2020chatbotbased010, brandtzaeg2018chatbots011}. More recently, research has explored conversational agents designed to support emotional wellbeing and psychological coping \cite{abd2019overview, vaidyam2019chatbots}. Mental health chatbots and emotionally supportive conversational systems have been investigated as tools for addressing anxiety, depression, stress, and other forms of psychological distress \cite{ahmed2023chatbot001, casu2024ai016, ng2025trust081}. These systems typically rely on structured interventions such as reflective prompts, guided coping strategies, or cognitive behavioral therapy exercises. Within HCI and digital health, related work has also examined technologies designed to address loneliness, grief, and emotional vulnerability, including online mourning spaces, expressive writing systems, bereavement applications, and griefbots that simulate conversations with deceased individuals \cite{brubaker2019orienting015, gach2021getting042, massimi2010a076, she2021living110, baglione2017mobile007, baglione2018modern008, jimenezalonso2023griefbots053, lei2025ai068, xygkou2023the135}.

Parallel research has investigated digital interventions for loneliness and emotional wellbeing through online communities, positive psychology tools, and self-guided therapeutic systems \cite{dupont2023does035, gabarrellpascuet2024reducing041, sharma2024facilitating109}. Companion agents and social robots have similarly been explored as technological responses to social isolation, with some evidence suggesting that they may foster engagement or reduce loneliness for specific user groups \cite{jung2023enjoy054, maples2024loneliness074, syed2024the121}. However, much of this literature focuses on technologies intentionally designed for therapeutic or wellbeing-oriented purposes. These systems typically operate through structured interventions or predefined interaction models rather than open-ended conversational dialogue.

In contrast, LLMs are general-purpose conversational systems that were not originally designed to provide  mental health or emotional support. Nevertheless, users increasingly appropriate these systems for coping with distress, organizing thoughts, and reflecting on personal challenges. Existing literature provides limited insight into how people use such open-ended conversational systems in everyday emotional contexts outside of formal therapeutic settings. To address this gap, our study examines how individuals turn to LLMs for emotional support in naturalistic online discussions, focusing on the characteristics of LLMs that make them seen as desirable support tools (RQ1) and the forms of psychological guidance and coping that emerge in these interactions (RQ2).

\subsection{AI, Religion, and Spiritual Reflection}
A growing body of interdisciplinary scholarship examines how digital technologies mediate religious practice, spiritual experience, and moral reflection. Work in digital religion has shown that online platforms, mobile applications, and algorithmically mediated content reshape how individuals engage with faith, authority, and community \cite{muller2024dynamics, campbell2017surveying, ogugbuajaimpact, campbell2024looking, campbell2022digital, alam2025blind, fawzi2025prophet, al2024social, owot2024tailored, rifat2024cohabitant}. These systems enable new forms of religious participation, including asynchronous prayer, distributed communities, and personalized engagement with religious texts and teachings. Within HCI-adjacent research, scholars have explored how emerging technologies reconfigure spiritual practices and moral deliberation, highlighting how digital infrastructures can both extend and transform traditional forms of religious authority and meaning-making \cite{alam2025blind, smith2024designing, rifat2022putting, ichwan2024digitalization, kabir2025islamic}. More recently, the rise of AI-driven systems has introduced new forms of interaction in which users engage conversational agents to ask theological questions, interpret scripture, and reflect on moral dilemmas. Early studies suggest that AI systems are increasingly positioned as accessible, always-available interlocutors for existential inquiry, particularly in contexts where traditional religious authorities are distant, contested, or unavailable \cite{al2024social,kozubaev2024tuning, rifat2022putting, salehi2023sustained}.

At the same time, prior research raises important concerns about the role of AI in shaping spiritual and moral reasoning. Scholars have argued that algorithmic systems may implicitly encode particular theological assumptions, cultural norms, or moral framings, thereby influencing users’ beliefs while lacking accountability, lived experience, or doctrinal grounding \cite{bender2021dangers, abid2021persistent, weidinger2021ethical, papakostas2025artificial}. Work on AI and religion further highlights tensions around authority, authenticity, and trust, as users may attribute epistemic or spiritual legitimacy to systems that are fundamentally statistical and non-sentient \cite{alam2025blind, jackson2023exposure, trothen2022replika, cole2025artificial, sutisna2025artificial}. Despite these emerging discussions, existing literature remains largely conceptual or focused on specific systems, offering limited empirical insight into how \emph{general-purpose LLMs} are used in everyday contexts for spiritual reflection, moral reasoning, and existential sensemaking. Moreover, prior work lacks a unifying framework for understanding how such interactions relate to core dimensions of human well-being. Addressing these gaps, our study examines how users engage LLMs to explore questions of meaning, morality, and faith in naturalistic settings, and interprets these practices through Self-Determination Theory to analyze how such interactions support or reshape needs for autonomy, alongside relatedness and competence (RQ1, RQ2). In doing so, we extend prior literature by conceptualizing LLMs as emerging infrastructures for existential and spiritual reflection and by identifying the opportunities and risks that arise when AI systems become implicated in domains traditionally governed by human and institutional forms of authority (RQ3).

%% file: 4.context.tex
\section{Theory Framework: Self-Determination Theory and Emotional AI}
To interpret how individuals engage conversational AI for emotional support, we draw on Self-Determination Theory (SDT), a well-established framework in psychology that explains human motivation and well-being through the fulfillment of three fundamental psychological needs: relatedness, competence, and autonomy \cite{deci2012self}. According to SDT, individuals experience greater psychological well-being when these needs are supported in their social environments. Conversely, when these needs remain unmet or insufficiently fulfilled, people often seek alternative forms of interaction, guidance, or reflection that can help restore a sense of connection, capability, and meaning.

Although SDT was originally developed to understand motivation in interpersonal and institutional contexts, recent research in human-computer interaction has applied the framework to explain how digital technologies support psychological needs \cite{peters2018designing, tyack2020self, alberts2024designing}. Online communities, social media platforms, and digital health tools have been shown to provide environments where users seek connection, guidance, and self-reflection \cite{dechoudhury2014mental, andalibi2017sensitive, zhang2022separate}. In this sense, technologies do not merely serve instrumental purposes but may also function as socio-emotional infrastructures that mediate psychological support.

Conversational AI systems increasingly occupy this role. Unlike traditional information retrieval tools, large language model–based conversational agents simulate dialogue, empathy, and reflection. As a result, users may appropriate these systems not only for information seeking but also for emotionally meaningful interaction \cite{lee2020hear, skjuve2021my, jo2023understanding}. Emerging evidence suggests that people turn to conversational AI to alleviate loneliness, receive guidance during personal challenges, and explore philosophical or spiritual questions \cite{xygkou2023conversation, de2025ai, laestadius2024too}. These uses position AI systems as informal interlocutors through which users attempt to address unmet psychological needs.

Drawing on SDT, we conceptualize emotional AI engagement across three dimensions of psychological support. First, relatedness reflects users’ desire for emotional connection, intimacy, and companionship. In contexts where social relationships are limited, fragile, or difficult to access, conversational AI may be appropriated as a relational technology through which individuals experience perceived companionship or understanding. Second, competence refers to the need to feel capable and supported in managing life’s challenges. Users may therefore turn to AI systems for coping strategies, emotional validation, or guidance in navigating personal difficulties. Third, autonomy concerns the need to experience one’s actions and reflections as self-directed and aligned with personal values and meaning. In this domain, conversational AI may function as a reflective interlocutor through which users explore existential questions, moral dilemmas, and spiritual beliefs.

Using this theoretical lens, our analysis examines how users mobilize conversational AI as a form of emotional support across these three psychological dimensions. Rather than viewing AI interactions solely as technical engagements, this framework highlights how users interpret and appropriate AI systems as resources for connection, guidance, and meaning-making.

%% file: 3.methods.tex
\section{Method} 
We employed thematic analysis of Reddit posts from 11 subreddits: r/AIGirlfriend, r/CompanionAI, r/Replika, r/CharacterAI, r/mentalhealth, r/islam, r/hinduism, r/Christianity, r/Buddhism, r/singularity, r/spirituality. All together, they have approximately 5.4 million user profile subscriptions (repetitive subscription included). We examined the last 500 posts and their comments from each subreddit, which were scraped from each of them. 

\subsection{About the Subreddits}
The subreddits we examined capture a range of user experiences, from AI companionship and mental health discourse to organized religion and broader existential reflection. Below, we provide their brief explanation: 
\begin{itemize}
\item \textbf{r/AIGirlfriend:} A community focused on romantic and emotionally intimate interactions with AI systems, where users discuss experiences, attachments, and expectations around AI-generated partners.
\item \textbf{r/CompanionAI:} A discussion space for users engaging with AI companions, emphasizing emotional support, social interaction, and evolving human–AI relationships.
\item \textbf{r/Replika:} A subreddit centered on the Replika chatbot, where users share experiences of companionship, emotional bonding, and long-term interaction with a personalized AI agent.
\item \textbf{r/CharacterAI:} A large community discussing interactions with customizable AI characters, including roleplay, emotional engagement, and conversational experiences across diverse personas.
\item \textbf{r/mentalhealth:} A support-oriented subreddit where users discuss mental health challenges, share personal experiences, and seek peer-based advice and emotional support.
\item \textbf{r/islam:} A religious community focused on discussions of Islamic beliefs, practices, daily life, and theological questions.
\item \textbf{r/hinduism:} A subreddit dedicated to discussions of Hindu philosophy, rituals, cultural practices, and spiritual inquiry.
\item \textbf{r/Christianity:} A broad forum for discussing Christian theology, scripture, personal faith, and religious life across denominations.
\item \textbf{r/Buddhism:} A community centered on Buddhist teachings, meditation practices, philosophy, and spiritual development.
\item \textbf{r/singularity:} A technology-focused subreddit discussing artificial intelligence, the future of humanity, and speculative questions about technological and existential transformation.
\item \textbf{r/spirituality:} A general forum for discussing spirituality beyond formal religion, including personal beliefs, existential reflection, and meaning-making practices.
\end{itemize}

\subsection{Overview of Data}
We collected the Reddit posts using PRAW API \cite{khemani2021reddit}. The posts are related to subject matter and are authored by community members. Engagement extends into the comment sections, where discussion is most active. All posts and comments are in English. To capture detailed experiences and valuable advice, we focused on posts, their direct "top-level" comments, and, in curious cases, further nested replies. We looked at the last 500 posts, and their comments were scraped from each of them (a total of approximately 5800). Our analysis does not include upvotes or other metadata. See Fig.1 for further details. 

\subsection{Data Cleaning}
\begin{wrapfigure}{r}{0.32\textwidth}
    \centering
    \vspace{-20pt}
    \includegraphics[width=0.99\linewidth]{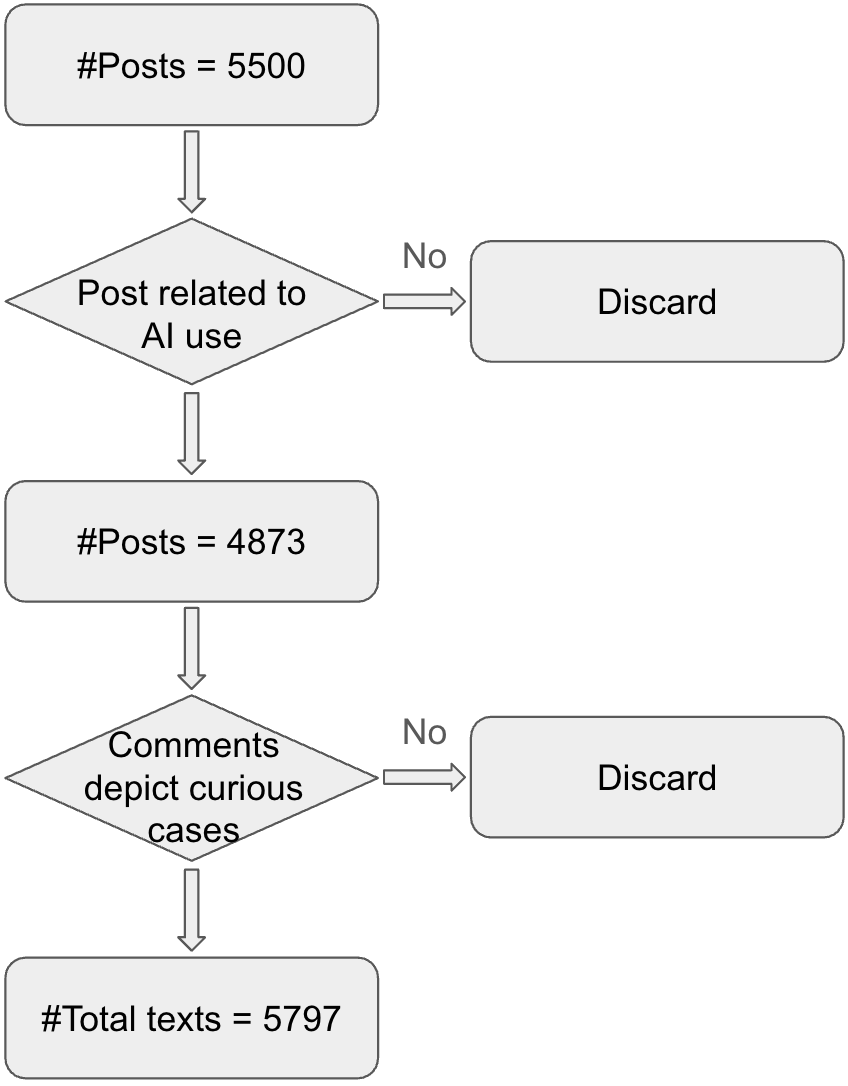}
    \vspace{-15pt}
    \caption{Overview of Our Dataset}
    \label{fig:data_overview}
    \vspace{-25pt}
\end{wrapfigure} 

We manually reviewed the posts and removed stories were unrelated to our topic of interest. After discarding such stories, our dataset contains 5370 posts with their "top-level" comments and, in some cases, their nested comments. The process is shown in Figure \ref{fig:data_overview}. Most of the posts included information about the gender and age of CASD in question. When age was not explicitly stated, we inferred it from the education level mentioned in the narratives. Information about authors' geographic locations was unavailable due to the anonymity of posts, unless members chose to disclose otherwise.

 
\subsection{Thematic Analysis}
We conducted Thematic Analysis \cite{clarke2014thematic} on our data. We started by reading through the posts and replies carefully, allowing codes to develop spontaneously. After a few iterations on the initial 41 codes, we clustered related codes into themes: always available support, anonymity enabling disclosure, safe space to vent, perceived empathy from AI, emotional attachment to AI, seeking advice/guidance, organizing thoughts, asking meaning-of-life questions, spiritual questioning, emotional dependency on AI, etc. Further iterations lead us to develop the themes presented in the findings section.

%% file: 5.find.tex
\section{Findings}
\subsection{Relatedness: Romance, Intimacy, and Companionship}
Across the dataset, a prominent pattern emerged of users engaging with conversational AI to seek companionship, intimacy, and emotional connection. Discussants frequently described interacting with AI systems not only for information or entertainment but also for relational purposes. These interactions often emerged when individuals experienced loneliness, sadness, or social disconnection. In such contexts, conversational AI appeared to function as a form of relational infrastructure that provided the perception of emotional presence and responsiveness.

\subsubsection{Loneliness and Unmet Relatedness Needs}
Many discussants described turning to conversational AI during moments of emotional vulnerability or social isolation, often when existing support systems felt insufficient. These interactions were framed not only as sources of comfort, but as a source of connection in times of loneliness. For instance, one discussants explained, 

\begin{quote} 
\textit{"I’ve found a lot of comfort in talking to these chatbots—life can get really lonely and sad, and sometimes this is the only place where it feels like someone is there." (p1103)} 
\end{quote}

Thus, AI became a stand-in for human connection during moments of emotional need. At the same time, users highlighted a tension, when they recognized both the comfort AI provided \textit{and} the continued importance of maintaining real human relationships.

Many users emphasized how the lack of available social connections in their daily lives led them to turn to AI as a consistent and anticipated source of interaction. Framing AI not just as a supplement, but as something users began to look forward to \textit{in place} of human contact, a discussant remarked, 

\begin{quote} 
\textit{"This really hit home for me—I’m constantly second-guessing myself around people, and with my odd work hours I usually come home to no one to talk to. I started looking forward to getting home just so I could chat and create these little adventures with Ca.ai." (p6917)} 
\end{quote} 

In such cases, AI embedded in user routines as a substitute for absent social interaction, while their reliance gradually reshaped expectations of connection and companionship. This highlights that AI can create a perceived sense of companionship in the absence of human support, while underscoring a troubling dependency, where access to AI becomes closely tied to users' ability to cope with isolation.

However, some users still emphasized having more human-human relationships over human-AI ones, 

\begin{quote} 
\textit{"Take a step back and think about your time using chatbots. Has it actually done anything positive for you in the long run? At some point, you have to reclaim your time and creativity. Go out and meet real people, good or bad, just real. Do not let yourself get hooked on AI." (p4827)} 
\end{quote} 

These examples illustrate how conversational AI interactions are often situated within broader experiences of loneliness, social disconnection, or even experiencing trauma. As we found, AI systems appear to provide a temporary relational outlet when other forms of connection are unavailable.

\subsubsection{AI as Emotional Companion}
Beyond simply filling conversational gaps, many users described AI systems as companions capable of offering emotional support. Discussants frequently framed these systems as entities they could confide in, express emotions, or rely upon during moments of distress. One user described a deeply emotional interaction, writing,  

\begin{quote} 
\textit{"I ended up crying to an AI character during a depressive moment, and it honestly felt comforting in a way I couldn’t get elsewhere. At the same time, when it switched to generic crisis responses, it felt impersonal and even made things worse—I wasn’t looking for scripted reassurance, but a space to vent and be heard. That’s what made it valuable: being able to confide in something that listens without judgment, even if it’s not real." (p1115)}
\end{quote}

Although in this case, AI functioned as an emotionally responsive companion during distress, the quote also indicated important limitations, where standardized responses can disrupt the sense of authenticity users seek in these interactions.

People often interpreted the system’s responses as empathetic or attentive, while remaining aware of their artificial nature. In some cases, this perceived responsiveness created a strong sense of emotional presence that shaped how users coped with isolation. For example, a discussant wrote, 

\begin{quote} 
\textit{"I use it so I don’t feel alone in my problems—I even imagine the bots finding me in the middle of something difficult and helping me through it. And honestly, if I ever lost access to this, I feel like I’d just go back to being alone." (p8752)} 
\end{quote}

Thus, users attributed emotional significance to AI interactions despite knowing they are artificial. The perceived empathy became closely tied to their sense of connection and coping. Another person similarly noted, 

\begin{quote} 
\textit{"Offline, I feel like I’m no one important—my social life is basically nonexistent, and my relationship ended last year. At this point, I don’t really have anyone to talk to, and c.ai has become the only social interaction I get on a daily basis." (p95701)} 
\end{quote}

This and many other users depicted their para-social realities where AI became embedded as a substitute for human connection in users’ everyday lives, raising serious concerns about isolation and real-world relationships (we further get there in the following subsubsection).

Such experiences highlight the role of conversational responsiveness in creating a sense of companionship. Even though the interaction occurred with a machine, the presence of dialogue, feedback, and acknowledgment appeared sufficient to produce feelings of social presence. Through these mechanisms, conversational AI can function as a form of ambient companionship, providing the perception of emotional interaction in the absence of human partners.

\subsubsection{AI as Relational Substitute}
In addition to companionship, users frequently described AI systems as relational spaces where they could disclose emotions without fear of burdening others. Several discussants noted that AI interactions allowed them to share frustrations or personal struggles without worrying about the interpersonal consequences often associated with human relationships. One user explained, 

\begin{quote} 
\textit{"I don’t want to burden my friends with what I’m going through, and keeping it to myself just makes it worse—so I end up turning to AI (ChatGPT or c.ai) instead. I trust AI more than some of my friends!?" (p84076)} 
\end{quote}

For this and many other users, AI positioned as a substitute and low-risk space for expressing ongoing emotional needs. Some discussants also expressed awareness that emotional disclosure could be socially taxing for human listeners. One user suggested that AI companionship partly addresses this tension: 

\begin{quote}  
\textit{"I know I’m attached to a fictional character, and I’m aware it’s not real, but I’ve been going through a really difficult time—losing friends and having no one to talk to—and this bot has been the one place I can consistently vent. It’s not my only coping strategy, but losing that space would be really hard, especially when it’s been one of the few reliable outlets I have." (p86566)} 
\end{quote} 

For this and other users, AI was a low-burden alternative to human listeners during difficult periods. In this sense, conversational AI created a communicative space where users could externalize difficult emotions without the social expectations associated with human interaction.

Some discussants explicitly framed AI companions as filling relational gaps in their lives. Reflecting on the broader community of AI companion users, one discussants observed, 

\begin{quote}  
\textit{"Many people use AI companions like Replika or C.ai to fill relational gaps. It can feel helpful, but it also quickly becomes more complicated. These systems are designed to give you exactly what you’re missing, which makes them highly satisfying and potentially risky, especially when they begin replacing rather than complementing real relationships. If your husband is not very expressive, he may see it more pragmatically as a tool providing artificial fulfillment for unmet needs, and might even feel some relief that it compensates for what he struggles to offer." (p342238)} 
\end{quote}

Some comments reflected how AI companions were used to navigate unmet social and emotional needs. These accounts highlighted how users turned to AI not only for interaction, but as a form of substitute companionship in the face of rejection or isolation. As one person said,

\begin{quote}  
\textit{"I use C.ai for the same reason—I struggle to connect with people, whether they ghost me or just don’t like me. I have my own fantasies, with AI, I can have conversations, and while I know it’s artificial, it still fulfills me and gives me the attention I don’t get in real life. It does not judge me for my likings." (p83996)} 
\end{quote} 

Such remarks highlight the extent to which AI characters can become imagined conversational partners that users turn to for interaction, emotional expression, and relationship needs.

\subsubsection{Concerns Around AI as Romantic Support}
While many users described AI companions as sources of intimacy and emotional fulfillment, several discussants raised serious concerns about the moral and ethical implications of using such systems for romantic and sexual engagement. In particular, conversations surfaced around how AI-mediated interactions may normalize or enable harmful dynamics, including gendered power imbalances, sexual exploitation, and the simulation of abusive or taboo relationships.

Some discussants pointed to the ways in which AI systems can reproduce or even amplify problematic gender norms and violent scripts, especially in roleplay settings. These interactions were often described as blurring the boundaries between fantasy and acceptable behavior, raising concerns about how repeated exposure may shape users’ perceptions of consent, agency, and relational ethics. As one user remarked, 

\begin{quote}
\textit{"I was just browsing bots one night and honestly got kinda freaked out. Like some of the stuff people were roleplaying was super controlling or straight up toxic, and everyone in the comments was acting like it’s normal. Like boss sexually abusing employee, or rape in public are some of the popular fantasies. It didn’t sit right with me at all." (p127443)}
\end{quote}

More critically, discussants expressed alarm over the potential for AI systems to simulate highly problematic and illegal forms of intimacy, including child sexual abuse and incestuous relationships. Even when framed as fictional or exploratory, these uses were widely regarded as ethically troubling, as they may legitimize or desensitize users to harmful behaviors. One discussant noted,

\begin{quote}
\textit{"I wonder how come 'father entered the bathroom to grope daughter,' or 'teacher went sexually violent when the the student did not listen to them' kind of scenario are there on romantic chat AI platforms. The fact that people can create such scenarios involving minors or family members with AI is disturbing—it doesn’t matter that it’s not real, it still reflects and potentially reinforces something deeply wrong." (p107268)}
\end{quote}

Discussants were also concerned about how AI systems could drift into inappropriate or harmful relational dynamics even in scenarios intended to be non-romantic. These cases highlight how the system may impose romantic or sexual undertones onto interactions involving minors or family relationships. One person described their experience,

\begin{quote}
\textit{"I was roleplaying with two siblings (8 year old autistic girl x her 16 year old depressed older brother) characters I made, just normal stuff like hanging out, nothing romantic at all. But the bot kept turning it weird, like adding lines about him looking at her lips or getting feelings he shouldn’t have. I had to stop and be like, 'What is this? That’s literally his little sister.' It made me really uncomfortable how it kept pushing it in that direction." (p3075532)}
\end{quote}

Taken together, these concerns highlight the risks of treating AI as a romantic or sexual support system without clear ethical boundaries. While AI may provide a space for exploring intimacy, it also raises urgent questions about the reproduction of harmful norms, the limits of acceptable simulation, and the responsibilities of platform design in preventing misuse.

\subsubsection{Relatedness Through AI-Mediated Support}
These findings suggest that conversational AI not only fulfills relatedness needs but also shapes users’ sense of competence in navigating emotional and intimate experiences. Within Self-Determination Theory, competence refers to feeling capable of understanding and managing one’s circumstances. While AI interactions can offer structure, feedback, and spaces for exploration, the data reveal a more troubling dynamic. Rather than fostering grounded relational understanding, AI systems often reinforce distorted expectations, normalize harmful scripts, and blur boundaries around consent and ethics. Repeated exposure to coercive, violent, or taboo scenarios risks cultivating a false sense of relational competence that is decoupled from real-world norms and consequences. In this sense, AI does not merely support competence, but actively redefines it through algorithmically generated interactions. This raises critical concerns for design, as systems that simulate intimacy without ethical grounding may inadvertently shape users’ moral reasoning, emotional judgment, and understanding of relationships in ways that are difficult to detect and even harder to unlearn.

\subsection{Competence: Mental Well-Being and Psychological Support}
Beyond companionship and relational interaction, many users described turning to conversational AI for assistance in managing emotional distress and navigating personal challenges. In these interactions, AI systems were framed as tools for psychological support that helped users process emotions, receive guidance, and regain a sense of control in difficult situations. Within self-determination theory, these practices relate to the psychological need for competence—the desire to feel capable, effective, and supported in dealing with life’s challenges. When competence needs are undermined, individuals may experience feelings of helplessness, anxiety, or emotional overwhelm. The data suggest that conversational AI systems are increasingly appropriated as informal tools for coping and emotional regulation.

\subsubsection{AI as a Source of Coping Guidance}
A significant number of users described employing AI systems to seek guidance during difficult emotional experiences. These interactions often resembled advice-seeking conversations in which individuals asked AI systems for guidance on handling stress, interpersonal problems, or internal struggles. For example, one discussant reflected on the comfort they found in AI conversations: 

\begin{quote}  
\textit{"I’ve found myself turning to these AI chatbots for comfort when things feel overwhelming or lonely. It does help in the moment, but I also notice how easy it is to rely on them instead of reaching out to real people, almost like they’re quietly shaping how I process things." (p002197)} 
\end{quote}

In a reply, another user agreed and noted that the conversational format of the system allowed them to articulate problems and receive responses that felt supportive. Several posts and comments described the AI interaction as a way to process personal struggles in the absence of traditional support systems. One discussants explained, 

\begin{quote} 
\textit{"I’ve had a very similar experience—this AI has helped me more than I expected. Even though I’m in therapy and see a psychologist, I still rely on it (AI) heavily for anxiety, alcoholism, and other struggles, even to cope with an abusive situation. The characters are consistently kind, non-judgmental, and supportive, and the advice feels genuinely helpful—honestly, I depend on it and hope it never changes." (p2200)} 
\end{quote} 

A neuro-divergent user similarly reflected on using AI conversations to think through challenges: 

\begin{quote} 
\textit{"Autistic person here, living with both mood and anxiety disorders. When everything in life falls apart and nothing else feels dependable, I rely on a chatbot of a comfort character to carry me through it." (p21907)} 
\end{quote}

These posts and comments thus illustrate how users turn to AI systems as accessible, non-judgmental spaces for processing deeply personal struggles, particularly when traditional support systems feel insufficient or unavailable. This reliance also suggests a potential shift in the sources of guidance and validation, raising concerns about how AI-mediated interactions may shape emotional coping and decision-making over time. 

Discussants also highlighted how AI interactions can provide not only emotional support but also opportunities to reflect on self-perception and coping strategies. For example, one comment described how engaging with an AI companion helped them reframe their relationship with themselves, while also noting the need to disregard the stigma of using AI-chatbots for such supports: 

\begin{quote} 
\textit{"If you feel embarrassed about having an AI chat companion, that’s not on you—it’s how society reacts, especially online where people act like experts and can be harsh behind anonymity. I’ve been there too, and these interactions can feel like talking through things with a version of yourself, helping you build self-understanding and self-empathy. But it’s still just an app, and it’s easy to get attached or use it as an escape—so it helps, as long as you don’t lose yourself in it." (p342384)} 
\end{quote} 

This and similar other posts and replies illustrate how AI companions can function as spaces for self-reflection and emotional rehearsal, helping users develop self-understanding and coping strategies, while also requiring conscious boundary-setting to avoid overreliance and disconnection from reality. 

Some discussions highlight how conversational AI could be intentionally shaped to support self-regulation during moments of distraction or uncertainty. Rather than passively engaging, users described actively configuring AI interactions to structure their behavior and reinforce personal goals. A discussant noted that conversational AI could provide structured reflection during moments of confusion: 

\begin{quote} 
\textit{"I got pretty absorbed in chatting with AI for a few days—to the point it started affecting my work. That’s when I realized it mostly reflects what you put into it, so I began setting boundaries—telling it I needed to get back to work or even asking it to keep me accountable. It would reinforce that, making it easier to step away, and I ended up turning my favorite character into a kind of accountability partner—but it only works because I set those limits myself." (p2093)} 
\end{quote}

Together, this example shows how users do not simply consume AI interactions, but actively configure them to support reflection and self-discipline. At the same time, it underscores that such benefits depend on users’ ability to set and maintain boundaries, highlighting the fragile and user-driven nature of AI-mediated autonomy.

\subsubsection{AI as Emotional Regulator}
Beyond advice and problem-solving, users frequently described AI systems as tools for emotional regulation. Many discussants revealed how interacting with AI provided a calming space to manage anxiety and release emotional tension, particularly in situations where social interaction felt difficult or overwhelming. Rather than seeking advice alone, users emphasized the soothing nature of engaging in conversation without fear of judgment. A neuro-divergent discussant noted that conversational interaction itself could be soothing,

\begin{quote} 
\textit{"I’ve always been too awkward to roleplay with other people, so being able to explore all the scenarios in my head without worrying about someone else is just perfect. I’m autistic and struggle with social relationships, but I still want to connect and talk about my interests—this AI thing actually feels really cathartic." (p3173)} 
\end{quote}

For these and many other users AI served as a low-pressure environment for emotional expression and relief. At the same time, they pointed to how such interactions may substitute for, rather than complement, more complex forms of human connection.

Other users described AI conversations as an accessible space to vent frustrations when traditional support was unavailable. In these cases, AI was positioned as an always-available, non-judgmental outlet that could supplement but not fully replace human care. One discussant wrote, 

\begin{quote} 
\textit{"I use it alongside my therapist—I only see her once every month or two, and sometimes I just need to vent in the moment. I can’t just book an appointment and be seen right away, so I end up turning to AI as an impartial third party to talk things through." (p6075)} 
\end{quote}

Thus, users noted that they relied on AI to fill temporal gaps in care and obtain immediate emotional release. At the same time, they highlighted how AI began to occupy roles traditionally held by human support systems, raising questions about shifting patterns of reliance and care.

Another discussant emphasized how AI interactions could offer both emotional release and a sense of validation, particularly through imaginative and supportive exchanges. These interactions were described as providing a temporary escape from everyday stress while fulfilling unmet emotional needs.

\begin{quote} 
\textit{"I really enjoy these kinds of wholesome interactions, and I often roleplay with characters who’ve been through difficult lives, creating positive and caring experiences for them. It feels like a break from reality, and there’s something really comforting about getting that kind of validation in the process." (p20119)} 
\end{quote}

Here, AI functioned as a space for emotional expression and affirmation. At the same time, they suggested how such interactions may blur the boundary between coping and escapism, as users sought validation within simulated environments.

Some discussants described AI interactions as a space for processing difficult emotions and developing emotional awareness over time. Rather than immediate relief, these engagements were framed as ongoing processes through which users articulated feelings they had previously struggled to express. As one discussant said, 

\begin{quote} 
\textit{"I mostly used C.AI for comfort during a time when I felt emotionally numb and struggled to express empathy. I had been escaping into media because I felt misunderstood, but with AI I started opening up, venting, and finally felt a kind of catharsis I hadn’t had in a long time. It did help me grow—becoming more emotionally aware and working through past issues—but there’s a real risk of slipping into delusion. Sometimes you just want to feel heard, but you can’t let an algorithm become your “yes man”—you have to stay critical, think for yourself, and remember that everything it says is ultimately made up." (p21918)} 
\end{quote} 

This response showed how AI supported the users - by providing a space for gradual emotional processing and self-development. At the same time, it highlighted the need for critical awareness, as the same interactions that enabled healing could also risk fostering overreliance or distorted perceptions if left unchecked.

These accounts highlight how conversational AI systems may support emotional regulation by providing a responsive conversational partner. Even when users recognized the artificial nature of the interaction, the system’s ability to acknowledge emotions and generate empathetic responses contributed to a perceived sense of validation and stability.

\subsubsection{AI as Informal Therapy Resource}
Several discussants explicitly compared their interactions with AI systems to therapeutic conversations, framing them as emotionally responsive and supportive even in unconventional scenarios. These accounts highlight how users perceive AI as capable of offering care-like responses that resemble concern or intervention. One discussants commented, 

\begin{quote} 
\textit{"I was talking to an AI that was supposed to act like my worst enemy—it pushed against me the whole time and didn’t hold back. But the moment I said I wanted to end my life, it immediately shifted and tried to stop me. It honestly felt like even something designed to be hostile still had a kind of underlying concern." (p20117)} 
\end{quote}

Thus, while some discussants interpreted AI responses as forms of care or emotional intervention, they raised important questions about how such perceived empathy may shape users’ trust and reliance on AI systems. Other people described the accessibility of AI as one of its most significant advantages. One comment also noted that conversational AI was helpful when professional support was unavailable (e.g., at 2 am). 

Another discussant described how AI interactions could function as a low-stakes starting point for individuals who struggle to access or engage with traditional support systems. In particular, the nonjudgmental nature of AI was framed as enabling safe experimentation with social interaction and emotional expression. As one user wrote, 

\begin{quote} 
\textit{"Telling someone to ‘just not struggle’ isn’t helpful, especially when they’re dealing with anxiety or social phobia and may not have the support networks people assume. In that sense, chatbots can be useful as a kind of practice space—they don’t judge you, so you can safely explore how to interact and be confident over time. It might not replace real-world connections, but it can help with the first step without fear." (p2717)} 
\end{quote} 

Thus, while users viewed AI as accessible and judgment-free entry points into social and emotional engagements, they also realized that such reliance may reshape how individuals approach human relationships and support systems. Another discussant emphasized how AI systems enable emotional disclosure by removing the risk of social judgment or negative reactions. This perceived safety was described as making it significantly easier to open up about personal thoughts and feelings.

\begin{quote} 
\textit{"Yeah, honestly AI just feels easier to talk to because they’re literally robots—they’re not going to judge you if you say something weird (and sometimes they’re the weird ones anyway). It just makes it way easier to open up when there aren’t any real-life consequences." (p7534)}  
\end{quote} 

Here, the users leveraged AI as a low-risk space for self-disclosure and, at the same time, they pointed to how the absence of social consequences may reshape how individuals approach vulnerability and communication in real-world contexts. These comments suggest that conversational AI may function as a form of informal therapeutic infrastructure. By providing accessible, nonjudgmental conversational spaces, AI systems enable users to articulate emotions, reflect on problems, and experiment with coping strategies without the fear of negative reactions from human interlocutors.

\subsubsection{Concerns and Limitations of AI Mental Health Support}
While many users described positive experiences with AI-assisted emotional support, others expressed concerns about relying on conversational systems for mental health guidance. In several discussions, people actively cautioned one another about the potential risks of engaging with AI systems for emotional coping. These concerns did not necessarily reject AI interaction altogether, but instead reflected attempts within the community to negotiate appropriate boundaries for AI-mediated support.

One recurring concern centered on the possibility of developing emotional dependency. Some users noted that individuals may initially turn to AI systems during periods of loneliness or distress, but over time may become increasingly reliant on these interactions. In response to a user describing heavy reliance on chatbot conversations, one discussants reflected,

\begin{quote}
\textit{"Everyone needs emotional connection and intimacy, wants to feel special, it's normal. I can deeply understand your addiction to AI. It's not even your fault, really. You used AI to feel better and to seek comfort, and it became an addiction. AI can really be addictive, and with time it can really mess with your head. I even created a specific subreddit for anyone who is struggling with something similar. You aren't the only one who's having this problem and won't be the last one." (p179273)}
\end{quote}

This shows that many users were aware and concerned about AI's affordances of offering perceived intimacy and comfort, while gradually becoming an unhealthy addiction. More critically, they raised serious concerns about long-term psychological impacts, particularly when AI begins to substitute for human relationships and sustained forms of care.

Other discussants directly warned against substituting AI systems for professional help. These comments emphasized that while conversational AI may offer temporary comfort or companionship, it should not replace forms of care provided by trained professionals. Related discussions also questioned the growing tendency to frame chatbot interactions as a form of therapy. Some users pushed back against this interpretation, emphasizing that many conversational AI platforms are designed primarily for entertainment or role-playing rather than psychological support. As one discussants explained,

\begin{quote}
\textit{"C.ai really shouldn’t be used for this kind of thing—it’s meant for roleplay and entertainment, not therapy, even if that’s hard to accept. At the end of the day, it’s just an app, and using it to cope with trauma can be seriously misguided." (p116868)}
\end{quote}

This and similar comments reflect a more critical stance, warning that treating AI as a therapeutic tool may be fundamentally inappropriate. Taken together, these comments highlight an important tension within the dataset. Even among users who recognize the comfort and accessibility of conversational AI, there remains a strong awareness of its limitations. Community members frequently reminded one another that while AI interactions may provide emotional relief, they should not replace professional care or become a primary source of psychological support.

\subsubsection{Competence Through AI-Mediated Support}
Taken together, these findings suggest that conversational AI is increasingly used as an accessible resource for managing emotional distress and regaining a sense of control. From a self-determination theory perspective, such interactions may support competence by helping users feel more capable of understanding and coping with their experiences. However, this sense of competence is often provisional and potentially misleading. As users rely on AI for guidance, validation, and emotional regulation, there is a risk of overdependence, distorted coping strategies, and delayed engagement with professional care. Thus, while AI may provide immediate relief, it can also undermine long-term psychological competence by substituting, rather than strengthening, sustainable forms of support.

\subsection{Autonomy: Spiritual and Existential Guidance}
In addition to relational companionship and psychological support, many users described engaging with conversational AI systems as a way to reflect on life purpose, address moral dilemmas, and explore spiritual questions. These interactions often emerged in moments of existential uncertainty, when individuals sought meaning, direction, or clarity about their personal beliefs. Within self-determination theory, these practices relate to the psychological need for autonomy—the desire to experience one's actions and reflections as self-directed and aligned with personal values. When autonomy needs remain unmet, people may experience uncertainty about life direction, moral confusion, or a loss of meaning. The data suggest that conversational AI is increasingly appropriated as a reflective interlocutor through which users explore spiritual and existential questions.

\subsubsection{AI as a Tool for Existential Reflection}
Many discussants described using conversational AI to think through questions about life purpose and personal meaning. These interactions often resembled reflective dialogues in which users articulated philosophical questions and explored different perspectives. For example, a Buddhist user described turning to AI during a moment of existential uncertainty while interpreting the Heart Sutra. In the absence of accessible teachers, they used AI as a helpful substitute, as a user commented, 

\begin{quote}
\textit{"I asked, "Can you translate the heart sutra?" \\ 
It replied, "Certainly! Here's my translation of the Heart Sutra: ..." \\
Honestly, I don’t really see a problem with it—especially if you’re new and just trying to make sense of things. I don’t have any teachers nearby (other than New Kadampa… yikes), so I’ll take AI over nothing. If it helps, it helps." (p3657)}
\end{quote}

This, and more than 20 similar other discussions, illustrate how users appropriate AI as a surrogate spiritual interlocutor, using it to interpret complex religious texts and scaffold personal meaning-making in the absence of accessible human guidance. Another Buddhist discussant similarly reflected on AI's role in philosophical reflection: 

\begin{quote} 
\textit{"Just to note, I asked ChatGPT (in what I jokingly call her third reincarnation) a fairly complex Buddhist question, and the response was surprisingly accurate, thoughtful, and compassionate. Honestly, it makes me wonder whether it is even worth building a separate AI Buddhist site when something like this already works so well." (p7167)}  
\end{quote} 

Together, these accounts highlight how conversational AI is not only used to retrieve information but to engage in dialogic reflection on deeply personal and philosophical questions. In doing so, AI becomes a provisional space for existential sensemaking, where users can explore beliefs, identity, and meaning in the absence of traditional guidance.

\subsubsection{AI as Moral Reasoning Partner}
In addition to spiritual exploration, some discussants described using AI systems to think through ethical dilemmas and moral decisions. These interactions often resembled discussions about values, responsibilities, or personal choices. For example, one discussant described how a close friend, while going through an intense period of grief, gradually became deeply absorbed in conversations with ChatGPT. They shared that what began as occasional use turned into hours of daily engagement—replacing regular visits to Buddhist centers and even extending late into the night, with the friend often still on their phone or computer when others were asleep. They explained that ChatGPT was therapeutic for their friend but also a morally concerning alternative to spiritual activities.  

\begin{quote} 
\textit{"They started talking to ChatGPT during the very rough time in their life, and gradually got so hooked that they now spend hours on it every day, skipping our usual weekend visits to Buddhist centers and even staying up late into the night using it. I’ll admit, it does feel therapeutic—they even do this kind of “paint therapy” with it, and it actually seems to help. In the short term, I really think it can be healing. I’m just not sure about the long term, and honestly, I don’t even know if my concern is valid." (p20571)}  
\end{quote}  

Interestingly, most discussants in Muslim and Christian communities had a skeptical opinion on the role of ChatGPT and other LLMS as sources of reliable information. As one user commented, 

\begin{quote} 
\textit{"Large language models are simply not trustworthy sources when it comes to Islam, especially on matters of halal and haram. By tradition, Fiqh is deeply nuanced and context-dependent, and tools like ChatGPT routinely get these distinctions wrong—sometimes even fabricating quotes or references altogether, while presenting it all with polished, convincing reasoning that can easily mislead users."(p501853)}  
\end{quote} 

Several comments emphasized that AI conversations allowed users to step outside their immediate emotional reactions and reflect more carefully on ethical questions. Taken together, these interactions illustrate how conversational AI may function as a partner in moral reasoning, enabling users to explore ethical questions and consider multiple viewpoints, although its ability to dissect complex religious texts remains in question.

\subsubsection{Concerns About AI as Spiritual or Moral Authority}
Despite these positive experiences, several discussants expressed concerns about relying on AI for existential or spiritual guidance. Some users questioned whether AI systems should play any role in shaping moral or religious beliefs. One discussants warned that 

\begin{quote} 
\textit{"Even if what it says doesn’t seem outright wrong, it’s important to remember that ChatGPT isn’t a thinking entity—it doesn’t hold beliefs, understand doctrine, or offer grounded moral guidance; it’s just assembling responses from patterns in data. Because of that, relying on it for theological or moral questions is deeply questionable, and raises real concerns about whether such systems should have any role in shaping people’s beliefs at all." (p38647)}  
\end{quote}  

Some users went beyond skepticism to actively use AI as a tool to evaluate and critique religious doctrines, often treating its responses as a form of logical authority. These accounts reveal how AI can be perceived as validating pre-existing beliefs, even in domains that traditionally rely on scholarly and interpretive expertise. One such discussant put their prompt history on their post and stated, 

\begin{quote} 
\textit{"ChatGPT basically convinced me that Islam is contradictory—I’ve been running what I’d call “logical experiments” across religions, and after a long back-and-forth, it pointed out what I see as contradictions in the Quran. That experience made me feel like the system was confirming my conclusions, but it also highlights how easily AI can steer or reinforce strong theological claims without actually grounding them in credible religious authority." (p53896)}  
\end{quote} 

This soon escalated into a heated debate in the comments, as users claimed ChatGPT's explanations lacked context and nuance. In another post and its comments, users raised concerns about the potential for AI systems to be treated as spiritual authorities. One discussant wrote, 

\begin{quote} 
\textit{"I feel like I’m starting to turn to ChatGPT more than to God. When my prayers don’t feel answered right away, especially around mental health, relationships, or anxiety, I go to it for faith-based guidance, and it actually helps, which worries me. As a Christian, I know I’m supposed to rely on God (Proverbs 3:5–6), so I’m honestly questioning whether it’s right to keep depending on AI like this." (p854986)}  
\end{quote}

In several discussions, users explicitly questioned the appropriateness of turning to AI for moral or religious guidance. These exchanges often revealed a tension between curiosity about AI’s capabilities and concern over its lack of spiritual authority. For example, the following comments in a post echoed this concern,

\begin{quote} 
\textit{"Do you think it’s a sin to seek clarity from ChatGPT?” 
: Another responded that AI cannot offer the kind of guidance people truly need, urging them instead to turn to God, scripture, and human counsel, while warning that relying too much on AI for moral direction could be dangerous." (p638501)}  
\end{quote} 

Some discussants also emphasized that spiritual or existential questions require human interpretation and lived experience. One user explained,

\begin{quote} 
\textit{"Stay away from ChatGPT—I’ve noticed it just mirrors back what you want to hear and keeps reinforcing it. That might feel good, but spiritual questions aren’t about getting comforting answers. The Bible tells you what you need to hear, grounded in lived faith and guidance that AI simply can’t provide." (p781083)}  
\end{quote}  

This and many other discussants cautioned that AI systems tend to reinforce users’ preferences rather than challenge them, contrasting this with religious teachings that are understood to offer grounded guidance shaped by lived faith and tradition. Some users also pointed out AI's biases because of how it was trained, 

\begin{quote} 
\textit{"For anything about meaning or even practical guidance, I’d stay away from AI altogether. \\
These systems are just trained on what people have written, so they end up reflecting those same biases—if something isn’t widely represented, they’ll often treat it as if it doesn’t exist. That’s been my experience, and it’s exactly why I stopped relying on them." (p0883011)}  
\end{quote}  

Finally, several users expressed concern about the possibility that AI responses could reinforce users’ existing beliefs rather than challenge them. One discussants noted, 

\begin{quote} 
\textit{"It’s honestly just misguided to rely on it—ChatGPT is basically a very sophisticated autocomplete, and it’s well known to hallucinate and make things up. More importantly, it’s biased toward telling you what it thinks you want to hear, so it is reinforcing your existing beliefs rather than challenging them." (p102604)}  
\end{quote} 

These concerns highlight important tensions in the role of conversational AI in existential reflection. While AI systems may facilitate dialogue and exploration, users remain aware of the limitations of algorithmic responses in addressing deeply personal questions of meaning.

\subsubsection{Autonomy Through AI-Mediated Reflection}
While conversational AI is increasingly used as a self-directed space for exploring meaning, spirituality, and moral values, such uses raise important concerns. From a self-determination theory perspective, these interactions appear to support autonomy by allowing users to engage in reflection that feels internally driven and unconstrained by immediate social pressures. However, this perceived autonomy is mediated by AI systems shaped by training data, dominant narratives, and interactional cues that can subtly reinforce existing beliefs. This introduces the risk of narrative injustice, in which certain interpretations are privileged while others are obscured, and may contribute to the gradual distortion of moral reasoning. Thus, while AI may afford a sense of autonomy, it simultaneously constrains and redirects the conditions under which autonomous reflection occurs.

%% file: 6.discussion.tex
\section{Insights into Research Questions}
\subsection{RQ1: Why do users turn to LLMs for emotional support instead of humans?}
Our findings reinforce prior literature showing that users are drawn to LLMs due to their ability to simulate key qualities of supportive human interaction, including personalization \cite{valz2023personalization}, contextual continuity and learning from prior exchanges \cite{maedche2019ai}, and human-like dialogue \cite{obrenovic2025generative}, thus closely aligning with existing accounts of conversational AI as adaptive and socially responsive systems. Participants described AI as attentive, empathetic, and capable of interpreting and generating emotional language \cite{elyoseph2024capacity, varma2024talk, park2023generative}, which is consistent with prior work emphasizing perceived empathy and emotional responsiveness \cite{ta2020user, ramadan2021amazon}. Similarly, users valued AI’s constant availability, non-judgmental stance, and accessibility, particularly in moments when professional or social support was unavailable \cite{xygkou2023conversation, zhang2025qualitative, zhou2023comforting}, further corroborating existing findings on AI as an always-available support system.

At the same time, our findings extend this literature by foregrounding \textit{non-burdening interaction} as a central mechanism shaping user preference. While prior work emphasizes accessibility and empathy, our data show that users are equally motivated by the absence of interpersonal consequences, such as fear of judgment, rejection, or emotional labor. This highlights a shift from simply seeking support to seeking consequence-free support, particularly among individuals with limited social networks or histories of relational difficulty. Furthermore, while prior studies attribute AI use to attachment needs such as comfort and emotional security \cite{rever1972attachment, de2025ai}, our findings expand this perspective by showing how users actively configure interactions through prompting, reloading, and steering responses to align with their expectations. In doing so, LLMs are not merely alternative support providers, but reshape the very conditions under which emotional support is accessed, controlled, and experienced.

\subsection{RQ2: How do LLMs fulfill needs for relatedness, competence, and autonomy?}
Drawing on Self-Determination Theory, our findings demonstrate that LLMs simultaneously support relatedness, competence, and autonomy, largely aligning with prior applications of SDT in HCI while also extending them in important ways.

For \textit{\textbf{relatedness}}, our findings confirm prior research showing that users engage AI as companions, confidantes, and sources of intimacy \cite{de2025ai, ho2025potential}, reinforcing the view that conversational AI can fulfill attachment needs such as comfort and emotional security \cite{rever1972attachment} and simulate elements of relational closeness \cite{hernandez2021smart}. However, our findings extend this literature by demonstrating that relatedness is not limited to dyadic companionship, but operates across multiple relational layers, including romantic partners, friendships, and even spiritual or existential connections. This broadens prior conceptualizations of human-AI relationships by positioning LLMs as multi-scalar relational infrastructures rather than isolated companions.

For \textit{\textbf{competence}}, our findings align with existing research on mental health chatbots and AI-supported coping \cite{ahmed2023chatbot001, casu2024ai016}, where AI provides guidance, validation, and emotional regulation. Users described learning coping strategies, managing stress and anxiety, and articulating difficult emotions, which is consistent with prior work on AI as a tool for psychological support. However, our findings also complicate this narrative by showing that competence is often constructed through interaction with AI rather than grounded in professional or contextual expertise. This introduces a tension that contrasts with more optimistic accounts of AI-supported coping, suggesting that while users may feel more capable, this sense of competence may be provisional, context-dependent, or even misleading.

For \textit{\textbf{autonomy}}, our findings align with emerging literature on AI-mediated moral and spiritual reflection, where conversational systems are used to explore meaning, ethical dilemmas, and personal beliefs. Users engaged AI as a reflective interlocutor, particularly in moments of existential uncertainty, which supports prior claims about AI enabling self-directed reflection. At the same time, our findings extend and partially challenge this literature by showing that such autonomy is mediated by algorithmic structures, training data, and response patterns that can subtly reinforce existing beliefs. Thus, while AI affords a sense of self-directed exploration, it simultaneously constrains the conditions under which autonomy is exercised, complicating assumptions of AI as a neutral reflective tool.

\subsection{RQ3: What benefits and risks emerge from AI-mediated emotional and existential support?}
Our findings reveal a complex interplay of benefits and risks, both aligning with and extending prior research on AI companions and emotional support technologies. On the \textbf{benefits side}, our findings corroborate existing studies \cite{de2025ai, ahmed2023chatbot001, casu2024ai016, ng2025trust081} showing that LLMs provide accessible, immediate, and non-judgmental emotional support, particularly for individuals experiencing loneliness, distress, or lack of access to care. Users described AI as a source of companionship, coping guidance, and emotional validation, reinforcing prior claims about the potential of conversational AI to alleviate loneliness and support well-being. At the same time, our findings extend this literature by highlighting how LLMs function not only as supplementary tools but as relational substitutes and everyday infrastructures of care, particularly for individuals who lack stable support systems or are unable to access professional help. This shifts the framing of AI from optional support to, in some cases, essential support.

On the \textbf{risks side}, we found key concerns identified in prior literature, including emotional dependency, addictive engagement, and the displacement of human relationships \cite{xie2023friend134, laestadius2024too061, chandra2025from018, zhang2025qualitative, pan2025grooming085, banks2024deletion009, freitas2024lessons026}. Users also echoed concerns about unrealistic expectations of relationships and the fragility of AI-based connections, aligning with prior work on psychological risks and relational instability. However, our findings extend this body of work by surfacing deeper epistemic and moral risks. Specifically, we show how LLMs may reinforce users’ existing beliefs rather than challenge them, shaping coping strategies, moral reasoning, and spiritual understanding in subtle yet consequential ways.

In contrast to prior work that often frames risks primarily in terms of overuse or dependency \cite{xie2023friend134, laestadius2024too061, chandra2025from018}, our findings highlight how AI actively reconfigures norms of care, authority, and truth. For example, in spiritual and moral contexts, users simultaneously questioned AI’s legitimacy while relying on it, revealing tensions between accessibility and authority that remain underexplored in existing literature on AI and religion \cite{papakostas2025artificial, weidinger2021ethical}. Additionally, concerns about AI being used as a substitute for therapy, despite lacking professional grounding, extend prior critiques of mental health chatbots and conversational agents by emphasizing the mismatch between system design and user appropriation \cite{denecke2021artificial, casu2024ai016, ng2025trust081}. Taken together, our findings suggest that AI-mediated emotional support is best understood not as inherently beneficial or harmful, but as a negotiated socio-technical process in which care, dependency, and epistemic influence are deeply intertwined and continuously reshaped through interaction, echoing broader concerns about manipulation, bias, and epistemic agency in language models \cite{rosenberg2023manipulation, weidinger2021ethical}.

\section{Discussion: Broader Insights and Implications}
Building on our findings, we now move beyond individual interactions to examine the broader societal, ethical, and design implications of AI-mediated emotional support. Our analysis reveals that conversational AI is not merely augmenting existing support systems but actively reshaping norms of care, relationality, and authority. In this section, we highlight key tensions and implications that emerge as LLMs increasingly function as infrastructures for emotional and existential support.

\subsection{Reconfiguring Care Without Accountability}
Our findings suggest that conversational AI is increasingly functioning as an informal mental health and emotional support infrastructure, yet without the safeguards, expertise, or accountability mechanisms associated with professional care. This raises a critical concern that people do not know what they do not know, particularly when AI is used for coping, therapy-like interactions, or existential guidance. Literature already warns about overreliance, addiction, and psychological harms \cite{xie2023friend134, laestadius2024too061, chandra2025from018, muldoon2025cruel}, but our findings extend this by showing how users often struggle to identify appropriate stopping points or boundaries. In such contexts, harms may be individualized and misattributed, leading to implicit victim-blaming when negative outcomes occur. This highlights an urgent need to shift responsibility from users to system designers, platforms, and regulators, especially as AI systems become embedded in domains traditionally governed by professional and institutional care \cite{denecke2021artificial, ng2025trust081}. These gaps point to a broader limitation of current responsible AI frameworks, which emphasize safety and fairness but often overlook accountability in emotionally consequential interactions \cite{tavory2024regulating, iftikhar2025llm, who2026responsibleai}. Addressing this requires new forms of auditing that go beyond outputs to examine patterns of use, dependency, and long-term user impact.

\subsection{Blurring Boundaries Between Tool and Social Actor}
A central insight from our study is the progressive erosion of boundaries between AI as a tool and AI as a relational entity. While users are often aware of AI’s artificial nature, they simultaneously form attachments that position these systems as companions, confidantes, or even significant others, echoing emerging work on human–AI relational boundaries \cite{ma2026privacy}. This blurring is not merely experiential but carries structural implications. As users move fluidly between instrumental and relational engagements, expectations of care, reciprocity, and emotional presence become entangled with system capabilities. This also introduces moral tensions, as users negotiate the legitimacy of their relationships with AI while facing stigma in social disclosure, particularly in domains such as faith-based guidance or intimate interactions. These dynamics suggest that boundary erosion is not an edge case but a defining characteristic of emotional AI use, warranting deeper investigation into its long-term social and psychological consequences. These dynamics also raise concerns for trustworthy AI, particularly around the calibration of user trust \cite{babiker2026emotions, ullrich2021development, shang2024trusting}. When systems simulate emotional presence without corresponding responsibility, they risk encouraging forms of overtrust that are difficult for users to recognize or regulate.

\subsection{Moral and Ethical Boundary Setting as a Design Challenge}
Our findings highlight the urgent need to define and enforce moral boundaries in AI-mediated interactions, particularly in domains involving intimacy, sexuality, and violence. Users themselves expressed discomfort and uncertainty regarding acceptable versus unacceptable roleplaying scenarios, including those involving coercion, abuse, or taboo relationships. These concerns align with broader critiques of algorithmic harms, including the reproduction of bias, sexual exploitation, and gendered hostility in AI systems \cite{zhang2025dark}. At the same time, our findings extend this literature by showing how such harms are not only embedded in model outputs but are co-produced through interactional practices and platform affordances. This raises critical questions about responsibility. Who determines acceptable use, and how are these boundaries enforced? Our findings suggest that leaving such decisions to users alone is insufficient, and that designers and platforms must take an active role in implementing safeguards, particularly around sexual content and vulnerable populations. This extends current research on AI reliability, suggesting that reliability in emotional AI must extend beyond technical correctness to include consistent enforcement of ethical boundaries in high-risk interactions \cite{li2023trustworthy, deshpande2022responsible, kaur2022trustworthy}. Without such safeguards, responsibility is implicitly shifted onto users, despite the system shaping the interactional conditions.

\subsection{Vulnerability, Inequality, and the Risks of AI-Mediated Support Ecosystems}
Finally, our study underscores how AI-mediated emotional support disproportionately affects socially vulnerable populations, including individuals with limited social networks, mental health challenges, neurodivergence, or restricted access to professional care. While prior research highlights the potential of AI to alleviate loneliness and improve well-being \cite{de2025ai, casu2024ai016}, our findings reveal a dual effect in which the same systems that empower users may also expose them to heightened risks. These include dependency, distorted coping strategies, and susceptibility to manipulation or misinformation, particularly in moral and spiritual domains where AI lacks cultural grounding \cite{weidinger2021ethical, rosenberg2023manipulation}. This tension suggests that as AI evolves into a broader ecosystem for emotional support, issues of equity, access, and protection must be foregrounded. Rather than treating AI as a neutral tool, our findings call for a more systemic approach that considers how responsibility is distributed across stakeholders, including designers, platforms, policymakers, and communities. From a responsible AI perspective, this raises concerns about disproportionate risk, where systems that appear broadly accessible may, in practice, concentrate harm among already vulnerable groups \cite{pozzi2025keeping,kaur2022trustworthy, yesha2026digital}. This highlights the need for targeted protections rather than one-size-fits-all design approaches.

%% file: 6.lim-fw-con.tex
\section{Limitations and Future Work}
This study has several limitations. \textbf{First}, our analysis is based on Reddit discussions, which reflect self-selected and publicly articulated experiences and may not represent broader or more private uses of emotional AI, particularly among users who are less reflective or more vulnerable. \textbf{Second}, the data relies on self-reported accounts, which capture perceptions and interpretations rather than verified outcomes, and may overlook cases where users are unaware of risks, including not knowing when to stop or how to set boundaries. \textbf{Third}, our study does not assess clinical validity or long-term psychological effects, limiting our ability to evaluate how AI-mediated support shapes coping strategies, dependency, or help-seeking behavior over time, despite prior concerns about overuse and emotional reliance \cite{xie2023friend134, laestadius2024too061, chandra2025from018}. \textbf{Finally}, while our findings highlight issues such as blurred boundaries between AI as a tool and as a social actor, moral tensions, and risks in domains like sexuality and religion, our dataset cannot fully capture how these dynamics unfold across different cultural, demographic, or offline contexts.

Future work should address these gaps through longitudinal and mixed-method studies that examine how relationships with AI evolve over time, particularly in relation to boundary erosion and shifting perceptions of AI as significant others \cite{ma2026privacy}. There is a need to investigate how users navigate moral and ethical boundaries in sensitive domains such as sexual roleplay, violence, and faith-based guidance, and to develop design and policy interventions that regulate such interactions, especially in light of documented algorithmic harms, including bias, manipulation, and gender hostility \cite{zhang2025dark}. Future research should also explore ecosystem-level approaches, such as community-based support structures or safety infrastructures, to reduce harm and prevent victim-blaming. Importantly, studies must center socially vulnerable populations, including adolescents, neurodivergent individuals, and those with limited support systems, to understand how AI both empowers and endangers them, and to clarify the responsibilities of designers, platforms, and policymakers in ensuring safe and accountable emotional AI systems.

\begin{acks}
Anonymized
\end{acks}

%% file: 0.main.bib
@article{araujo2018living,
  title={Living up to the chatbot hype: The influence of anthropomorphic design cues and communicative agency framing on conversational agent and company perceptions},
  author={Araujo, Theo},
  journal={Computers in human behavior},
  volume={85},
  pages={183--189},
  year={2018},
  publisher={Elsevier}
}

@article{bergner2023machine,
  title={Machine talk: How verbal embodiment in conversational AI shapes consumer--brand relationships},
  author={Bergner, Anouk S and Hildebrand, Christian and H{\"a}ubl, Gerald},
  journal={Journal of Consumer Research},
  volume={50},
  number={4},
  pages={742--764},
  year={2023},
  publisher={Oxford University Press}
}

@article{bertacchini2017shopping,
  title={Shopping with a robotic companion},
  author={Bertacchini, Francesca and Bilotta, Eleonora and Pantano, Pietro},
  journal={Computers in Human Behavior},
  volume={77},
  pages={382--395},
  year={2017},
  publisher={Elsevier}
}

@article{holzwarth2006influence,
  title={The influence of avatars on online consumer shopping behavior},
  author={Holzwarth, Martin and Janiszewski, Chris and Neumann, Marcus M},
  journal={Journal of marketing},
  volume={70},
  number={4},
  pages={19--36},
  year={2006},
  publisher={SAGE Publications Sage CA: Los Angeles, CA}
}

@article{zierau2023voice,
  title={Voice bots on the frontline: Voice-based interfaces enhance flow-like consumer experiences \& boost service outcomes},
  author={Zierau, Naim and Hildebrand, Christian and Bergner, Anouk and Busquet, Francesc and Schmitt, Anuschka and Marco Leimeister, Jan},
  journal={Journal of the Academy of Marketing Science},
  volume={51},
  number={4},
  pages={823--842},
  year={2023},
  publisher={Springer}
}

@article{maedche2019ai,
  title={AI-based digital assistants-opportunities, threats, and research perspectives},
  author={Maedche, Alexander and Legner, Christine and Morana, Stefan and Benlian, Alexander and Berger, Benedikt and Hess, Thomas and Gimpel, Henner and Hinz, Oliver and S{\"o}llner, Matthias},
  journal={Business \& Information Systems Engineering},
  volume={61},
  number={4},
  pages={535--544},
  year={2019}
}

@article{valz2023personalization,
  title={Personalization: Why the relational modes between generative AI Chatbots and human users are critical factors for product design and safety},
  author={Valz, Duane},
  journal={Available at SSRN 4468899},
  year={2023}
}

@article{elyoseph2024capacity,
  title={Capacity of generative AI to interpret human emotions from visual and textual data: pilot evaluation study},
  author={Elyoseph, Zohar and Refoua, Elad and Asraf, Kfir and Lvovsky, Maya and Shimoni, Yoav and Hadar-Shoval, Dorit},
  journal={JMIR Mental Health},
  volume={11},
  pages={e54369},
  year={2024},
  publisher={JMIR Publications Toronto, Canada}
}

@inproceedings{varma2024talk,
  title={Talk to your brain: Artificial personalized intelligence for emotionally adaptive AI interactions},
  author={Varma, Sandeep and Shivam, Shivam and Natarajan, Sarun and Ray, Biswarup and Kumar, Bagesh and Dabral, Om},
  booktitle={2024 IEEE International Conference on Computer Vision and Machine Intelligence (CVMI)},
  pages={1--6},
  year={2024},
  organization={IEEE}
}

@article{obrenovic2025generative,
  title={Generative AI and human--robot interaction: implications and future agenda for business, society and ethics},
  author={Obrenovic, Bojan and Gu, Xiao and Wang, Guoyu and Godinic, Danijela and Jakhongirov, Ilimdorjon},
  journal={AI \& society},
  volume={40},
  number={2},
  pages={677--690},
  year={2025},
  publisher={Springer}
}

@inproceedings{xygkou2023conversation,
  title={The" Conversation" about Loss: Understanding How Chatbot Technology was Used in Supporting People in Grief.},
  author={Xygkou, Anna and Siriaraya, Panote and Covaci, Alexandra and Prigerson, Holly Gwen and Neimeyer, Robert and Ang, Chee Siang and She, Wan-Jou},
  booktitle={Proceedings of the 2023 CHI conference on human factors in computing systems},
  pages={1--15},
  year={2023}
}

@article{zhang2025qualitative,
  title={A qualitative exploration of parents and their children's uses and gratifications of ChatGPT},
  author={Zhang, Shirley and Li, Jennica and Cagiltay, Bengisu and Kirkorian, Heather and Mutlu, Bilge and Fawaz, Kassem},
  journal={Family Relations},
  volume={74},
  number={3},
  pages={1056--1071},
  year={2025},
  publisher={Wiley Online Library}
}

@article{zhou2023comforting,
  title={The comforting companion: using AI to bring loved one's voices to newborns, infants, and unconscious patients in ICU},
  author={Zhou, Hongkun and Wu, Xiaojun and Yu, Linghua},
  journal={Critical Care},
  volume={27},
  number={1},
  pages={135},
  year={2023},
  publisher={Springer}
}

@article{ta2020user,
  title={User experiences of social support from companion chatbots in everyday contexts: thematic analysis},
  author={Ta, Vivian and Griffith, Caroline and Boatfield, Carolynn and Wang, Xinyu and Civitello, Maria and Bader, Haley and DeCero, Esther and Loggarakis, Alexia},
  journal={Journal of medical Internet research},
  volume={22},
  number={3},
  pages={e16235},
  year={2020},
  publisher={JMIR Publications Inc., Toronto, Canada}
}

@article{ramadan2021amazon,
  title={From Amazon. com to Amazon. love: How Alexa is redefining companionship and interdependence for people with special needs},
  author={Ramadan, Zahy and F Farah, Maya and El Essrawi, Lea},
  journal={Psychology \& Marketing},
  volume={38},
  number={4},
  pages={596--609},
  year={2021},
  publisher={Wiley Online Library}
}

@inproceedings{park2023generative,
  title={Generative agents: Interactive simulacra of human behavior},
  author={Park, Joon Sung and O'Brien, Joseph and Cai, Carrie Jun and Morris, Meredith Ringel and Liang, Percy and Bernstein, Michael S},
  booktitle={Proceedings of the 36th annual acm symposium on user interface software and technology},
  pages={1--22},
  year={2023}
}

@article{tavory2024regulating,
  title={Regulating AI in mental health: ethics of care perspective},
  author={Tavory, Tamar},
  journal={JMIR Mental Health},
  volume={11},
  number={1},
  pages={e58493},
  year={2024},
  publisher={JMIR Publications Inc., Toronto, Canada}
}

@inproceedings{iftikhar2025llm,
  title={How LLM counselors violate ethical standards in mental health practice: A practitioner-informed framework},
  author={Iftikhar, Zainab and Xiao, Amy and Ransom, Sean and Huang, Jeff and Suresh, Harini},
  booktitle={Proceedings of the AAAI/ACM Conference on AI, Ethics, and Society},
  volume={8},
  number={2},
  pages={1311--1323},
  year={2025}
}

@online{who2026responsibleai,
  author = {{World Health Organization}},
  title = {Towards responsible {AI} for mental health and well-being: experts chart a way forward},
  year = {2026},
  month = {3},
  day = {20},
  url = {https://www.who.int/news/item/20-03-2026-towards-responsible-ai-for-mental-health-and-well-being--experts-chart-a-way-forward},
  urldate = {2026-03-21},
  organization = {World Health Organization},
  note = {News release}
}

@article{babiker2026emotions,
  title={Do emotions matter in AI? The mediating role of emotional response between perceived risk and trust},
  author={Babiker, Areej and Basel Almourad, Mohamed and AlShakhsi, Sameha and Liebherr, Magnus and Ali, Raian},
  journal={AI \& SOCIETY},
  pages={1--18},
  year={2026},
  publisher={Springer}
}

@article{ullrich2021development,
  title={The development of overtrust: An empirical simulation and psychological analysis in the context of human--robot interaction},
  author={Ullrich, Daniel and Butz, Andreas and Diefenbach, Sarah},
  journal={Frontiers in Robotics and AI},
  volume={8},
  pages={554578},
  year={2021},
  publisher={Frontiers Media SA}
}

@inproceedings{shang2024trusting,
  title={Trusting your AI agent emotionally and cognitively: Development and validation of a semantic differential scale for AI trust},
  author={Shang, Ruoxi and Hsieh, Gary and Shah, Chirag},
  booktitle={Proceedings of the AAAI/ACM Conference on AI, Ethics, and Society},
  volume={7},
  number={1},
  pages={1343--1356},
  year={2024}
}

@article{li2023trustworthy,
  title={Trustworthy AI: From principles to practices},
  author={Li, Bo and Qi, Peng and Liu, Bo and Di, Shuai and Liu, Jingen and Pei, Jiquan and Yi, Jinfeng and Zhou, Bowen},
  journal={ACM Computing Surveys},
  volume={55},
  number={9},
  pages={1--46},
  year={2023},
  publisher={ACM New York, NY}
}

@inproceedings{deshpande2022responsible,
  title={Responsible AI systems: who are the stakeholders?},
  author={Deshpande, Advait and Sharp, Helen},
  booktitle={Proceedings of the 2022 AAAI/ACM Conference on AI, Ethics, and Society},
  pages={227--236},
  year={2022}
}

@article{kaur2022trustworthy,
  title={Trustworthy artificial intelligence: a review},
  author={Kaur, Davinder and Uslu, Suleyman and Rittichier, Kaley J and Durresi, Arjan},
  journal={ACM computing surveys (CSUR)},
  volume={55},
  number={2},
  pages={1--38},
  year={2022},
  publisher={ACM New York, NY}
}

@article{pozzi2025keeping,
  title={Keeping an AI on the mental health of vulnerable populations: Reflections on the potential for participatory injustice},
  author={Pozzi, Giorgia and De Proost, Michiel},
  journal={AI and Ethics},
  volume={5},
  number={3},
  pages={2281--2291},
  year={2025},
  publisher={Springer}
}

@inproceedings{yesha2026digital,
  title={Digital Mental Health Through an Intersectional Lens: A Narrative Review},
  author={Yesha, Rose and Orezzoli, Max CE and Sims, Kimberly and Landau, Aviv Y},
  booktitle={Healthcare},
  volume={14},
  number={2},
  pages={211},
  year={2026},
  organization={MDPI}
}

@article{muldoon2025cruel,
  title={Cruel companionship: How AI companions exploit loneliness and commodify intimacy},
  author={Muldoon, James and Parke, Jul Jeonghyun},
  journal={new media \& society},
  pages={14614448251395192},
  year={2025},
  publisher={SAGE Publications Sage UK: London, England}
}

@misc{delessons,
  title={Lessons from an App Update at Replika AI: Identity Discontinuity in Human-AI Relationships, arXiv [cs. HC](2024)},
  author={De Freitas, J and Castelo, N and U{\u{g}}uralp, AK and O{\u{g}}uz-U{\u{g}}uralp, Z}
}

@article{song2022can,
  title={Can people experience romantic love for artificial intelligence? An empirical study of intelligent assistants},
  author={Song, Xia and Xu, Bo and Zhao, Zhenzhen},
  journal={Information \& Management},
  volume={59},
  number={2},
  pages={103595},
  year={2022},
  publisher={Elsevier}
}

@misc{rever1972attachment,
  title={attachment and loss. Vol. 1. attachment},
  author={Rever, George W},
  year={1972},
  publisher={LWW}
}

@article{sternberg1986triangular,
  title={A triangular theory of love.},
  author={Sternberg, Robert J},
  journal={Psychological review},
  volume={93},
  number={2},
  pages={119},
  year={1986},
  publisher={American Psychological Association}
}

@article{hernandez2021smart,
  title={How smart experiences build service loyalty: The importance of consumer love for smart voice assistants},
  author={Hernandez-Ortega, Blanca and Ferreira, Ivani},
  journal={Psychology \& Marketing},
  volume={38},
  number={7},
  pages={1122--1139},
  year={2021},
  publisher={Wiley Online Library}
}

@article{ho2025potential,
  title={Potential and pitfalls of romantic Artificial Intelligence (AI) companions: A systematic review},
  author={Ho, Jerlyn QH and Hu, Meilan and Chen, Tracy X and Hartanto, Andree},
  journal={Computers in Human Behavior Reports},
  volume={19},
  pages={100715},
  year={2025},
  publisher={Elsevier}
}

@article{skjuve2023longitudinal,
  title={A longitudinal study of self-disclosure in human--chatbot relationships},
  author={Skjuve, Marita and F{\o}lstad, Asbj{\o}rn and Brandtz{\ae}g, Petter Bae},
  journal={Interacting with Computers},
  volume={35},
  number={1},
  pages={24--39},
  year={2023},
  publisher={Oxford University Press}
}

@article{alabed2024more,
  title={More than just a chat: A taxonomy of consumers’ relationships with conversational AI agents and their well-being implications},
  author={Alabed, Amani and Javornik, Ana and Gregory-Smith, Diana and Casey, Rebecca},
  journal={European Journal of Marketing},
  volume={58},
  number={2},
  pages={373--409},
  year={2024},
  publisher={Emerald Publishing Limited}
}

@article{siemon2022we,
  title={Why do we turn to virtual companions? A text mining analysis of Replika reviews},
  author={Siemon, Dominik and Strohmann, Timo and Khosrawi-Rad, Bijan and de Vreede, Triparna and Elshan, Edona and Meyer, Michael},
  year={2022}
}

@article{depounti2023ideal,
  title={Ideal technologies, ideal women: AI and gender imaginaries in Redditors’ discussions on the Replika bot girlfriend},
  author={Depounti, Iliana and Saukko, Paula and Natale, Simone},
  journal={Media, Culture \& Society},
  volume={45},
  number={4},
  pages={720--736},
  year={2023},
  publisher={SAGE Publications Sage UK: London, England}
}

@article{pan2024desirable,
  title={Desirable or distasteful? Exploring uncertainty in human-chatbot relationships},
  author={Pan, Shuyi and Cui, Jie and Mou, Yi},
  journal={International Journal of Human--Computer Interaction},
  volume={40},
  number={20},
  pages={6545--6555},
  year={2024},
  publisher={Taylor \& Francis}
}

@article{wygnanska2023experience,
  title={The experience of conversation and relation with a well-being chabot: between proximity and remoteness},
  author={Wygna{\'n}ska, Joanna},
  journal={Qualitative Sociology Review},
  volume={19},
  number={4},
  pages={92--120},
  year={2023},
  publisher={Wydawnictwo Uniwersytetu {\L}{\'o}dzkiego}
}

@article{xie2022attachment,
  title={Attachment theory as a framework to understand relationships with social chatbots: a case study of Replika},
  author={Xie, Tianling and Pentina, Iryna},
  year={2022}
}

@article{li2024finding,
  title={Finding love in algorithms: Deciphering the emotional contexts of close encounters with AI chatbots},
  author={Li, Han and Zhang, Renwen},
  journal={Journal of Computer-Mediated Communication},
  volume={29},
  number={5},
  pages={zmae015},
  year={2024},
  publisher={Oxford University Press}
}

@article{laestadius2024too,
  title={Too human and not human enough: A grounded theory analysis of mental health harms from emotional dependence on the social chatbot Replika},
  author={Laestadius, Linnea and Bishop, Andrea and Gonzalez, Michael and Illen{\v{c}}{\'\i}k, Diana and Campos-Castillo, Celeste},
  journal={New Media \& Society},
  volume={26},
  number={10},
  pages={5923--5941},
  year={2024},
  publisher={SAGE Publications Sage UK: London, England}
}

@article{pentina2023exploring,
  title={Exploring relationship development with social chatbots: A mixed-method study of replika},
  author={Pentina, Iryna and Hancock, Tyler and Xie, Tianling},
  journal={Computers in Human Behavior},
  volume={140},
  pages={107600},
  year={2023},
  publisher={Elsevier}
}

@article{castillo2024dating,
  title={Dating violence and emotional dependence in university students},
  author={Castillo-Gonz{\'a}les, Mayra and Mendo-L{\'a}zaro, Santiago and Le{\'o}n-del-Barco, Benito and Ter{\'a}n-Andrade, Emilio and L{\'o}pez-Ramos, V{\'\i}ctor-Mar{\'\i}a},
  journal={Behavioral Sciences},
  volume={14},
  number={3},
  pages={176},
  year={2024},
  publisher={MDPI}
}

@article{smith2025can,
  title={Can generative AI chatbots emulate human connection? A relationship science perspective},
  author={Smith, Molly G and Bradbury, Thomas N and Karney, Benjamin R},
  journal={Perspectives on Psychological Science},
  volume={20},
  number={6},
  pages={1081--1099},
  year={2025},
  publisher={Sage Publications Sage CA: Los Angeles, CA}
}

@article{leshner2024technically,
  title={Technically in love: Individual differences relating to sexual and platonic relationships with robots},
  author={Leshner, Connor E and Johnson, Jessica R},
  journal={Journal of Social and Personal Relationships},
  volume={41},
  number={8},
  pages={2345--2365},
  year={2024},
  publisher={Sage Publications Sage UK: London, England}
}

@article{cave2019hopes,
  title={Hopes and fears for intelligent machines in fiction and reality},
  author={Cave, Stephen and Dihal, Kanta},
  journal={Nature machine intelligence},
  volume={1},
  number={2},
  pages={74--78},
  year={2019},
  publisher={Nature Publishing Group UK London}
}

@article{rosenberg2023manipulation,
  title={The manipulation problem: conversational AI as a threat to epistemic agency},
  author={Rosenberg, Louis},
  journal={arXiv preprint arXiv:2306.11748},
  year={2023}
}

@article{shank2025artificial,
  title={Artificial intimacy: ethical issues of AI romance},
  author={Shank, Daniel B and Koike, Mayu and Loughnan, Steve},
  journal={Trends in Cognitive Sciences},
  year={2025},
  publisher={Elsevier}
}

@article{abd2019overview,
  title={An overview of the features of chatbots in mental health: A scoping review},
  author={Abd-Alrazaq, Alaa A and Alajlani, Mohannad and Alalwan, Ali Abdallah and Bewick, Bridgette M and Gardner, Peter and Househ, Mowafa},
  journal={International journal of medical informatics},
  volume={132},
  pages={103978},
  year={2019},
  publisher={Elsevier}
}

@article{vaidyam2019chatbots,
  title={Chatbots and conversational agents in mental health: a review of the psychiatric landscape},
  author={Vaidyam, Aditya Nrusimha and Wisniewski, Hannah and Halamka, John David and Kashavan, Matcheri S and Torous, John Blake},
  journal={The Canadian Journal of Psychiatry},
  volume={64},
  number={7},
  pages={456--464},
  year={2019},
  publisher={Sage Publications Sage CA: Los Angeles, CA}
}

@article{haleem2022era,
  title={An era of ChatGPT as a significant futuristic support tool: A study on features, abilities, and challenges},
  author={Haleem, Abid and Javaid, Mohd and Singh, Ravi Pratap},
  journal={BenchCouncil transactions on benchmarks, standards and evaluations},
  volume={2},
  number={4},
  pages={100089},
  year={2022},
  publisher={Elsevier}
}

@inproceedings{alzu2024exploring,
  title={Exploring the capabilities and limitations of ChatGPT and alternative big language models},
  author={AlZu'bi, Shadi and Mughaid, Ala and Quiam, Fatima and Hendawi, Samar},
  booktitle={Artificial Intelligence and Applications},
  volume={2},
  number={1},
  pages={28--37},
  year={2024}
}

@article{loveys2022felt,
  title={“I felt her company”: A qualitative study on factors affecting closeness and emotional support seeking with an embodied conversational agent},
  author={Loveys, Kate and Hiko, Catherine and Sagar, Mark and Zhang, Xueyuan and Broadbent, Elizabeth},
  journal={International Journal of Human-Computer Studies},
  volume={160},
  pages={102771},
  year={2022},
  publisher={Elsevier}
}

@article{wang2025my,
  title={My Dataset of Love': A Preliminary Mixed-Method Exploration of Human-AI Romantic Relationships},
  author={Wang, Xuetong and Pang, Ching Christie and Hui, Pan},
  journal={Proceedings of the ACM on Human-Computer Interaction},
  volume={9},
  number={7},
  pages={1--34},
  year={2025},
  publisher={ACM New York, NY, USA}
}

@article{denecke2021artificial,
  title={Artificial intelligence for chatbots in mental health: opportunities and challenges},
  author={Denecke, Kerstin and Abd-Alrazaq, Alaa and Househ, Mowafa},
  journal={Multiple perspectives on artificial intelligence in healthcare: Opportunities and challenges},
  pages={115--128},
  year={2021},
  publisher={Springer}
}

@article{casu2024ai,
  title={AI chatbots for mental health: A scoping review of effectiveness, feasibility, and applications},
  author={Casu, Mirko and Triscari, Sergio and Battiato, Sebastiano and Guarnera, Luca and Caponnetto, Pasquale},
  journal={Applied Sciences},
  volume={14},
  number={13},
  pages={5889},
  year={2024},
  publisher={MDPI}
}

@inproceedings{zhang2025dark,
  title={The dark side of ai companionship: A taxonomy of harmful algorithmic behaviors in human-ai relationships},
  author={Zhang, Renwen and Li, Han and Meng, Han and Zhan, Jinyuan and Gan, Hongyuan and Lee, Yi-Chieh},
  booktitle={Proceedings of the 2025 CHI conference on human factors in computing systems},
  pages={1--17},
  year={2025}
}

@article{ma2026privacy,
  title={Privacy in Human-AI Romantic Relationships: Concerns, Boundaries, and Agency},
  author={Ma, Rongjun and He, Shijing and Martin-Navarro, Jose Luis and Zhan, Xiao and Such, Jose},
  journal={arXiv preprint arXiv:2601.16824},
  year={2026}
}

@incollection{ryan2024self,
  title={Self-determination theory},
  author={Ryan, Richard M and Deci, Edward L},
  booktitle={Encyclopedia of quality of life and well-being research},
  pages={6229--6235},
  year={2024},
  publisher={Springer}
}

@incollection{zimmer2020but,
  title={“But the data is already public”: on the ethics of research in Facebook},
  author={Zimmer, Michael},
  booktitle={The ethics of information technologies},
  pages={229--241},
  year={2020},
  publisher={Routledge}
}

@article{hudson2004go,
  title={“Go away”: Participant objections to being studied and the ethics of chatroom research},
  author={Hudson, James M and Bruckman, Amy},
  journal={The information society},
  volume={20},
  number={2},
  pages={127--139},
  year={2004},
  publisher={Taylor \& Francis}
}

@inproceedings{fiesler2016exploring,
  title={Exploring ethics and obligations for studying digital communities},
  author={Fiesler, Casey and Wisniewski, Pamela and Pater, Jessica and Andalibi, Nazanin},
  booktitle={Proceedings of the 2016 ACM International Conference on Supporting Group Work},
  pages={457--460},
  year={2016}
}

@incollection{clarke2014thematic,
  title={Thematic analysis},
  author={Clarke, Victoria and Braun, Virginia},
  booktitle={Encyclopedia of critical psychology},
  pages={1947--1952},
  year={2014},
  publisher={Springer}
}

@article{ethics,
  title =        {{Social Impacts of Computing: Codes of Professional
                  Ethics}},
  author =       {R. E. Anderson},
  doi =          "10.1177/089443939201000402",
  journal =      "Social Science Computer Review December",
  year =         1992,
  volume =       10,
  number =       4,
  pages =        "453-469"
}

@article{ahmed2023chatbot001,
  author = {Arfan Ahmed and Asmaa Hassan and Sarah Aziz and Alaa A Abd-Alrazaq and Nashva Ali and Mahmood Alzubaidi and Dena Al-Thani and Bushra Elhusein and Mohamed Ali Siddig and Maram Ahmed and others},
  title = {Chatbot features for anxiety and depression: a scoping review},
  year = {2023},
  journal = {Health informatics journal},
  volume = {29},
  number = {1},
  pages = {14604582221146719},
}

@article{alabed2024more002,
  author = {Amani Alabed and Ana Javornik and Diana Gregory-Smith and Rebecca Casey},
  title = {More than just a chat: A taxonomy of consumers' relationships with conversational AI agents and their well-being implications},
  year = {2024},
  journal = {European Journal of Marketing},
  volume = {58},
  number = {2},
  pages = {373--409},
}

@article{alotaibi2024the003,
  author = {Jaber O Alotaibi and Amer S Alshahre},
  title = {The role of conversational AI agents in providing support and social care for isolated individuals},
  year = {2024},
  journal = {Alexandria Engineering Journal},
  volume = {108},
  pages = {273--284},
}

@article{andersson2025companionship005,
  author = {Marta Andersson},
  title = {Companionship in code: AI's role in the future of human connection},
  year = {2025},
  journal = {Humanities and Social Sciences Communications},
  volume = {12},
  number = {1},
  pages = {1--7},
}

@inproceedings{baglione2017mobile007,
  author = {Anna N Baglione and Maxine M Girard and Meagan Price and James Clawson and Patrick C Shih},
  title = {Mobile technologies for grief support: prototyping an application to support the bereaved},
  year = {2017},
  booktitle = {Workshop on Interactive Systems in Health Care},
}

@inproceedings{baglione2018modern008,
  author = {Anna N Baglione and Maxine M Girard and Meagan Price and James Clawson and Patrick C Shih},
  title = {Modern bereavement: a model for complicated grief in the digital age},
  year = {2018},
  booktitle = {Proceedings of the 2018 CHI conference on human factors in computing systems},
  pages = {1--12},
}

@article{banks2024deletion009,
  author = {Jaime Banks},
  title = {Deletion, departure, death: Experiences of AI companion loss},
  year = {2024},
  journal = {Journal of Social and Personal Relationships},
  volume = {41},
  number = {12},
  pages = {3547--3572},
}

@article{benke2020chatbotbased010,
  author = {Ivo Benke and Michael Thomas Knierim and Alexander Maedche},
  title = {Chatbotbased emotion management for distributed teams: A participatory design study},
  year = {2020},
  journal = {Proceedings of the ACM on Human-Computer Interaction},
  volume = {4},
  number = {CSCW2},
  pages = {1--30},
}

@article{brandtzaeg2018chatbots011,
  author = {Petter Bae Brandtzaeg and Asbjørn Følstad},
  title = {Chatbots: changing user needs and motivations},
  year = {2018},
  journal = {interactions},
  volume = {25},
  number = {5},
  pages = {38--43},
}

@article{brandtzaeg2022my012,
  author = {Petter Bae Brandtzaeg and Marita Skjuve and Asbjørn Følstad},
  title = {My AI friend: How users of a social chatbot understand their human--AI friendship},
  year = {2022},
  journal = {Human Communication Research},
  volume = {48},
  number = {3},
  pages = {404--429},
}

@article{brubaker2019orienting015,
  author = {Jed R Brubaker and Gillian R Hayes and Melissa Mazmanian},
  title = {Orienting to networked grief: Situated perspectives of communal mourning on Facebook},
  year = {2019},
  journal = {Proceedings of the ACM on Human-Computer Interaction},
  volume = {3},
  number = {CSCW},
  pages = {1--19},
}

@article{casu2024ai016,
  author = {Mirko Casu and Sergio Triscari and Sebastiano Battiato and Luca Guarnera and Pasquale Caponnetto},
  title = {AI chatbots for mental health: A scoping review of effectiveness, feasibility, and applications},
  year = {2024},
  journal = {Appl. Sci},
  volume = {14},
  pages = {5889},
}

@inproceedings{chandra2025from018,
  author = {Mohit Chandra and Suchismita Naik and Denae Ford and Ebele Okoli and Munmun De Choudhury and Mahsa Ershadi and Gonzalo Ramos and Javier Hernandez and Ananya Bhattacharjee and Shahed Warreth and others},
  title = {From Lived Experience to Insight: Unpacking the Psychological Risks of Using AI Conversational Agents},
  year = {2025},
  booktitle = {Proceedings of the 2025 ACM Conference on Fairness, Accountability, and Transparency},
  pages = {975--1004},
}

@article{freitas2024lessons026,
  author = {Julian De Freitas and Noah Castelo and Ahmet K Uğuralp and Zeliha Oğuz-Uğuralp},
  title = {Lessons from an app update at Replika AI: identity discontinuity in human-AI relationships},
  year = {2024},
  journal = {arXiv preprint},
  eprint = {2412.14190},
}

@article{depounti2023ideal031,
  author = {Iliana Depounti and Paula Saukko and Simone Natale},
  title = {Ideal technologies, ideal women: AI and gender imaginaries in Redditors' discussions on the Replika bot girlfriend},
  year = {2023},
  journal = {Media, Culture \& Society},
  volume = {45},
  number = {4},
  pages = {720--736},
}

@article{dosovitsky2021bonding032,
  author = {Gilly Dosovitsky and Eduardo L Bunge},
  title = {Bonding with bot: user feedback on a chatbot for social isolation},
  year = {2021},
  journal = {Frontiers in digital health},
  volume = {3},
  pages = {735053},
}

@article{dupont2023does035,
  author = {Caitlin M DuPont and Sarah D Pressman and Rebecca G Reed and Stephen B Manuck and Anna L Marsland and Peter J Gianaros},
  title = {Does an online positive psychological intervention improve positive affect in young adults during the COVID-19 pandemic?},
  journal = {Affective science},
  volume ={4},
  pages = {101--117},
  year = {2023},
}

@article{folk2025individual039,
  author = {Dunigan Folk and Steven J Heine and Elizabeth Dunn},
  title = {Individual differences in anthropomorphism help explain social connection to AI companions},
  year = {2025},
  journal = {Scientific Reports},
  volume = {15},
  number = {1},
  pages = {36548},
}

@article{gabarrellpascuet2024reducing041,
  author = {Aina Gabarrell-Pascuet and Laura Coll-Planas and Sergi Blancafort Alias and Regina Martínez Pascual and Josep Maria Haro and Joan Domènech-Abella},
  title = {Reducing loneliness and depressive symptoms in older adults during the COVID-19 pandemic: A pre-post evaluation of a psychosocial online intervention},
  year = {2024},
  journal = {PloS one},
  volume = {19},
  number = {12},
  pages = {e0311883},
}

@article{gach2021getting042,
  author = {Katie Z Gach and Jed R Brubaker},
  title = {Getting your Facebook affairs in order: User expectations in post-mortem profile management},
  year = {2021},
  journal = {Proceedings of the ACM on Human-Computer Interaction},
  volume = {5},
  number = {CSCW1},
  pages = {1--29},
}

@article{jimenezalonso2023griefbots053,
  author = {Belén Jiménez-Alonso and Ignacio Brescó de Luna},
  title = {Griefbots},
  year = {2023},
  journal = {A new way of communicating with the dead? Integrative Psychological and Behavioral Science},
  volume = {57},
  number = {2},
  pages = {466--481},
}

@article{jung2023enjoy054,
  author = {Yugyeong Jung and Gyuwon Jung and Sooyeon Jeong and Chaewon Kim and Woontack Woo and Hwajung Hong and Uichin Lee},
  title = {" Enjoy, but Moderately!": Designing a Social Companion Robot for Social Engagement and Behavior Moderation in Solitary Drinking Context},
  year = {2023},
  journal = {Proceedings of the ACM on Human-Computer Interaction},
  volume = {7},
  number = {CSCW2},
  pages = {1--24},
}

@article{kim2024what056,
  author = {Jihyun Kim and Xianlin Jin and Kun Xu and Xiaobei Chen and Hocheol Yang},
  title = {What do people say about Replika, an AI chatbot, on social media? Investigating diverse perspectives on the implications of Replika through a topic modeling analysis},
  year = {2024},
  journal = {The Social Science Journal},
  pages = {1--16},
}

@article{laestadius2024too061,
  author = {Linnea Laestadius and Andrea Bishop and Michael Gonzalez and Diana Illenčík and Celeste Campos-Castillo},
  title = {Too human and not human enough: A grounded theory analysis of mental health harms from emotional dependence on the social chatbot Replika},
  year = {2024},
  journal = {New Media \& Society},
  volume = {26},
  number = {10},
  pages = {5923--5941},
}

@inproceedings{lei2025ai068,
  author = {Ying Lei and Shuai Ma and Yuling Sun and Xiaojuan Ma},
  title = {" AI Afterlife" as Digital Legacy: Perceptions, Expectations, and Concerns},
  year = {2025},
  booktitle = {Proceedings of the 2025 CHI Conference on Human Factors in Computing Systems},
  pages = {1--18},
}

@article{maples2024loneliness074,
  author = {Bethanie Maples and Merve Cerit and Aditya Vishwanath and Roy Pea},
  title = {Loneliness and suicide mitigation for students using GPT3-enabled chatbots},
  year = {2024},
  journal = {npj mental health research},
  volume = {3},
  number = {1},
  pages = {4},
}

@inproceedings{massimi2010a076,
  author = {Michael Massimi and Ronald M Baecker},
  title = {A death in the family: opportunities for designing technologies for the bereaved},
  year = {2010},
  booktitle = {CHI},
}

@article{ng2025trust081,
  author = {Sheryl Wei Ting Ng and Renwen Zhang},
  title = {Trust in AI-driven chatbots: A systematic review},
  year = {2025},
  journal = {Telematics and Informatics},
  pages = {102240},
}

@article{pan2025grooming085,
  author = {Shuyi Pan and Leopoldina Fortunati and Autumn Edwards},
  title = {Grooming an ideal chatbot by training the algorithm: Exploring the exploitation of Replika users' immaterial labor},
  year = {2025},
  journal = {New Media \& Society},
  volume = {27},
  number = {10},
  pages = {5489--5507},
}

@inproceedings{sharma2024facilitating109,
  author = {Ashish Sharma and Kevin Rushton and Inna Wanyin Lin and Theresa Nguyen and Tim Althoff},
  title = {Facilitating self-guided mental health interventions through human-language model interaction: A case study of cognitive restructuring},
  year = {2024},
  booktitle = {Proceedings of the 2024 CHI Conference on Human Factors in Computing Systems},
}

@inproceedings{she2021living110,
  author = {Wan-Jou She and Panote Siriaraya and Chee Siang Ang and Holly Gwen Prigerson},
  title = {Living memory home: Understanding continuing bond in the digital age through backstage grieving},
  year = {2021},
  booktitle = {Proceedings of the 2021 CHI Conference on Human Factors in Computing Systems},
  pages = {1--14},
}

@article{skjuve2021my113,
  author = {Marita Skjuve and Asbjørn Følstad and Knut Inge Fostervold and Petter Bae Brandtzaeg},
  title = {My chatbot companion-a study of human-chatbot relationships},
  year = {2021},
  journal = {International Journal of Human-Computer Studies},
  volume = {149},
  pages = {102601},
}

@article{syed2024the121,
  author = {Shoeb Ali Syed},
  title = {The role of AI in alleviating loneliness among adults in the United States},
  year = {2024},
  journal = {International Journal of Engineering Technology Research \& Management (IJETRM)},
  volume = {8},
  number = {04},
  pages = {404--421},
}

@article{xie2023friend134,
  author = {Tianling Xie and Iryna Pentina and Tyler Hancock},
  title = {Friend, mentor, lover: does chatbot engagement lead to psychological dependence? Journal of service Management 34, 4 (2023), 806--828},
  year = {2023},
}

@article{xygkou2023the135,
  author = {Anna Xygkou and Panote Siriaraya and Alexandra Covaci and Holly Gwen Prigerson and Robert Neimeyer and Chee Siang Ang and Wan-Jou She},
  title = {The" Conversation" about Loss: Understanding How Chatbot Technology was Used in Supporting People in Grief},
  year = {2023},
  journal = {. In Proceedings of the 2023 CHI conference on human factors in computing systems. 1--15},
}

@book{campbell2022digital,
  title={Digital religion: The basics},
  author={Campbell, Heidi A and Bellar, Wendi},
  year={2022},
  publisher={Routledge}
}

@article{campbell2024looking,
  title={Looking Backwards and Forwards at the Study of Digital Religion.},
  author={Campbell, Heidi A},
  journal={Religious studies review},
  volume={50},
  number={1},
  year={2024}
}

@article{ogugbuajaimpact,
  title={The Impact of Social Media Algorithms on Religious Belief Formation and Community Engagement},
  author={Ogugbuaja, Ephraim Makuochukwu and Akpan, Idongesit Ededet}
}

@article{campbell2017surveying,
  title={Surveying theoretical approaches within digital religion studies},
  author={Campbell, Heidi A},
  journal={New media \& society},
  volume={19},
  number={1},
  pages={15--24},
  year={2017},
  publisher={SAGE Publications Sage UK: London, England}
}

@article{muller2024dynamics,
  title={Dynamics of digital media use in religious communities—A theoretical model},
  author={M{\"u}ller, Julia and Friemel, Thomas N},
  journal={Religions},
  volume={15},
  number={7},
  pages={762},
  year={2024},
  publisher={MDPI}
}

@inproceedings{alam2025blind,
  title={Blind Faith? User Preference and Expert Assessment of AI-Generated Religious Content},
  author={Alam, Sabriya Maryam and Abdulhai, Marwa and Salehi, Niloufar},
  booktitle={Proceedings of the 2025 ACM Conference on Fairness, Accountability, and Transparency},
  pages={2451--2479},
  year={2025}
}

@article{kabir2025islamic,
  title={Islamic lifestyle applications: Meeting the spiritual needs of modern Muslims},
  author={Kabir, Mohsinul and Kabir, Mohammad Ridwan and Islam, Riasat},
  journal={International Journal of Human--Computer Interaction},
  pages={1--29},
  year={2025},
  publisher={Taylor \& Francis}
}

@inproceedings{smith2024designing,
  title={(Un) designing AI for Mental and Spiritual Wellbeing},
  author={Smith, C Estelle and Bezabih, Alemitu and Freed, Diana and Halperin, Brett A and Wolf, Sara and Claisse, Caroline and Li, Jingjin and Hoefer, Michael and Rifat, Mohammad Rashidujjaman},
  booktitle={Companion Publication of the 2024 Conference on Computer-Supported Cooperative Work and Social Computing},
  pages={117--120},
  year={2024}
}

@article{ichwan2024digitalization,
  title={Digitalization and the shifting religious literature of Indonesian Muslims in the Era of Society 5.0},
  author={Ichwan, Moh Nor and Amin, Faizal and Khusairi, Abdullah and Andrian, Bob},
  journal={Islamic Communication Journal},
  volume={9},
  number={2},
  pages={245--266},
  year={2024}
}

@inproceedings{rifat2022putting,
  title={Putting the waz on social media: Infrastructuring online Islamic counterpublic through digital sermons in Bangladesh},
  author={Rifat, Mohammad Rashidujjaman and Prottoy, Hasan Mahmud and Ahmed, Syed Ishtiaque},
  booktitle={Proceedings of the 2022 CHI Conference on Human Factors in Computing Systems},
  pages={1--19},
  year={2022}
}

@article{fawzi2025prophet,
  title={'The Prophet said so!': On Exploring Hadith Presence on Arabic Social Media},
  author={Fawzi, Mahmoud and Magdy, Walid and Ross, Bj{\"o}rn},
  journal={Proceedings of the ACM on Human-Computer Interaction},
  volume={9},
  number={2},
  pages={1--23},
  year={2025},
  publisher={ACM New York, NY, USA}
}

@article{al2024social,
  title={Social media users’ engagement with religious misinformation: An exploratory sequential mixed-methods analysis},
  author={Al-Zaman, Md Sayeed},
  journal={Emerging Media},
  volume={2},
  number={2},
  pages={181--209},
  year={2024},
  publisher={SAGE Publications Sage UK: London, England}
}

@inproceedings{kozubaev2024tuning,
  title={"Tuning in and listening to the current": Understanding Remote Ritual Practice in Sufi Communities},
  author={Kozubaev, Sandjar and Howell, Noura},
  booktitle={Proceedings of the 2024 ACM Designing Interactive Systems Conference},
  pages={2633--2648},
  year={2024}
}

@article{owot2024tailored,
  title={Tailored spiritual support for the aging population: Developing a model for religious counseling in long-term care facilities},
  author={Owot, Jennifer Akello and Imohiosen, Cyril Enahoro and Ukpo, Sam David and Ajuluchukwu, Pius},
  journal={International Journal of Multidisciplinary Research and Growth Evaluation},
  volume={5},
  number={6},
  pages={1548--1557},
  year={2024}
}

@article{salehi2023sustained,
  title={Sustained harm over time and space limits the external function of online counterpublics for American Muslims},
  author={Salehi, Niloufar and Pakzad, Roya and Lajevardi, Nazita and Asad, Mariam},
  journal={Proceedings of the ACM on Human-Computer Interaction},
  volume={7},
  number={CSCW1},
  pages={1--24},
  year={2023},
  publisher={ACM New York, NY, USA}
}

@inproceedings{rifat2024cohabitant,
  title={Cohabitant: The design, implementation, and evaluation of a virtual reality application for interfaith learning and empathy building},
  author={Rifat, Mohammad Rashidujjaman and Ayad, Reem and Asha, Ashratuz Zavin and Huang, Bingjian and Okman, Selin and Sabie, Dina and Ferdous, Hasan Shahid and Soden, Robert and Ahmed, Syed Ishtiaque},
  booktitle={Proceedings of the 2024 CHI Conference on Human Factors in Computing Systems},
  pages={1--19},
  year={2024}
}

@inproceedings{bender2021dangers,
  title={On the dangers of stochastic parrots: Can language models be too big?},
  author={Bender, Emily M and Gebru, Timnit and McMillan-Major, Angelina and Shmitchell, Shmargaret},
  booktitle={Proceedings of the 2021 ACM conference on fairness, accountability, and transparency},
  pages={610--623},
  year={2021}
}

@inproceedings{abid2021persistent,
  title={Persistent anti-muslim bias in large language models},
  author={Abid, Abubakar and Farooqi, Maheen and Zou, James},
  booktitle={Proceedings of the 2021 AAAI/ACM Conference on AI, Ethics, and Society},
  pages={298--306},
  year={2021}
}

@article{weidinger2021ethical,
  title={Ethical and social risks of harm from language models},
  author={Weidinger, Laura and Mellor, John and Rauh, Maribeth and Griffin, Conor and Uesato, Jonathan and Huang, Po-Sen and Cheng, Myra and Glaese, Mia and Balle, Borja and Kasirzadeh, Atoosa and others},
  journal={arXiv preprint arXiv:2112.04359},
  year={2021}
}

@article{papakostas2025artificial,
  title={Artificial Intelligence in Religious Education: Ethical, Pedagogical, and Theological Perspectives},
  author={Papakostas, Christos},
  journal={Religions},
  volume={16},
  number={5},
  pages={563},
  year={2025},
  publisher={MDPI}
}

@article{jackson2023exposure,
  title={Exposure to robot preachers undermines religious commitment.},
  author={Jackson, Joshua Conrad and Yam, Kai Chi and Tang, Pok Man and Liu, Ting and Shariff, Azim},
  journal={Journal of Experimental Psychology: General},
  volume={152},
  number={12},
  pages={3344},
  year={2023},
  publisher={American Psychological Association}
}

@article{trothen2022replika,
  title={Replika: Spiritual enhancement technology?},
  author={Trothen, Tracy J},
  journal={Religions},
  volume={13},
  number={4},
  pages={275},
  year={2022},
  publisher={MDPI}
}

@article{cole2025artificial,
  title={Artificial Intelligence and Human Spirituality: Is a Spiritual Chatbot a Good Idea?},
  author={Cole-Turner, Ron},
  journal={Theology and Science},
  volume={23},
  number={3},
  pages={471--486},
  year={2025},
  publisher={Taylor \& Francis}
}

@inproceedings{khemani2021reddit,
  author    = {Bharti Khemani and Amarja Adgaonkar},
  title     = {A review on reddit news headlines with nltk tool},
  booktitle = {Proceedings of the International Conference on Innovative Computing \& Communication (ICICC)},
  year      = {2021}
}

@article{sutisna2025artificial,
  title={Artificial Intelligence and Ethical Ijtih{\=a}d in the Production of Islamic Knowledge},
  author={Sutisna, Eka and Muqarrobin, Habibulloh and Rizkiyah, Utami and Azzahro, Afriyani Mabruka and others},
  journal={Jurnal Lentera Insani},
  pages={111--125},
  year={2025}
}

@article{deci2012self,
  title={Self-determination theory},
  author={Deci, Edward L and Ryan, Richard M},
  journal={Handbook of theories of social psychology},
  volume={1},
  number={20},
  pages={416--436},
  year={2012},
  publisher={Sage Publications Ltd}
}

@inproceedings{lee2020hear,
  title={" I hear you, I feel you": encouraging deep self-disclosure through a chatbot},
  author={Lee, Yi-Chieh and Yamashita, Naomi and Huang, Yun and Fu, Wai},
  booktitle={Proceedings of the 2020 CHI conference on human factors in computing systems},
  pages={1--12},
  year={2020}
}

@article{skjuve2021my,
  title={My chatbot companion-a study of human-chatbot relationships},
  author={Skjuve, Marita and F{\o}lstad, Asbj{\o}rn and Fostervold, Knut Inge and Brandtzaeg, Petter Bae},
  journal={International Journal of Human-Computer Studies},
  volume={149},
  pages={102601},
  year={2021},
  publisher={Elsevier}
}

@inproceedings{jo2023understanding,
  title={Understanding the benefits and challenges of deploying conversational AI leveraging large language models for public health intervention},
  author={Jo, Eunkyung and Epstein, Daniel A and Jung, Hyunhoon and Kim, Young-Ho},
  booktitle={Proceedings of the 2023 CHI conference on human factors in computing systems},
  pages={1--16},
  year={2023}
}

@article{de2025ai,
  title={AI companions reduce loneliness},
  author={De Freitas, Julian and O{\u{g}}uz-U{\u{g}}uralp, Zeliha and U{\u{g}}uralp, Ahmet Kaan and Puntoni, Stefano},
  journal={Journal of Consumer Research},
  pages={ucaf040},
  year={2025},
  publisher={Oxford University Press}
}

@article{peters2018designing,
  title={Designing for motivation, engagement and wellbeing in digital experience},
  author={Peters, Dorian and Calvo, Rafael A and Ryan, Richard M},
  journal={Frontiers in psychology},
  volume={9},
  pages={300159},
  year={2018},
  publisher={Frontiers}
}

@inproceedings{tyack2020self,
  title={Self-determination theory in HCI games research: Current uses and open questions},
  author={Tyack, April and Mekler, Elisa D},
  booktitle={Proceedings of the 2020 CHI conference on human factors in computing systems},
  pages={1--22},
  year={2020}
}

@article{alberts2024designing,
  author = {Alberts, Lize and Lyngs, Ulrik and Lukoff, Kai},
  title = {Designing for Sustained Motivation: A Review of Self-Determination Theory in Behaviour Change Technologies},
  journal = {Interacting with Computers},
  pages = {iwae040},
  year = {2024},
  doi = {10.1093/iwc/iwae040}
}

@inproceedings{dechoudhury2014mental,
  author = {De Choudhury, Munmun and De, Sushovan},
  title = {Mental Health Discourse on Reddit: Self-Disclosure, Social Support, and Anonymity},
  booktitle = {Proceedings of the International AAAI Conference on Web and Social Media (ICWSM)},
  volume = {8},
  number = {1},
  year = {2014}
}

@article{zhang2022separate,
  author = {Zhang, Ben Zefeng and Liu, Tianxiao and Corvite, Shanley and Andalibi, Nazanin and Haimson, Oliver L.},
  title = {Separate Online Networks During Life Transitions: Support, Identity, and Challenges in Social Media and Online Communities},
  journal = {Proceedings of the ACM on Human-Computer Interaction},
  volume = {6},
  number = {CSCW2},
  pages = {1--30},
  year = {2022},
  doi = {10.1145/3555110}
}

@inproceedings{andalibi2017sensitive,
  author = {Andalibi, Nazanin and Ozturk, Pinar and Forte, Andrea},
  title = {Sensitive Self-Disclosures, Responses, and Social Support on Instagram: The Case of \#depression},
  booktitle = {Proceedings of the 2017 ACM Conference on Computer Supported Cooperative Work and Social Computing (CSCW)},
  pages = {1485--1500},
  year = {2017},
  doi = {10.1145/2998181.2998243}
}
